\documentstyle[12pt,psfig,epsfig]{article}

\newcommand{\resection}[1]{\setcounter{equation}{0}\section{#1}}

\newcommand{\EQ}{\begin{equation}}
\newcommand{\EN}{\end{equation}}
\newcommand{\bea}{\begin{eqnarray}}
\newcommand{\eea}{\end{eqnarray}}
\newcommand{\nn}{\nonumber}
\newcommand{\no}{\noindent}
\newcommand{\hs}{\hspace{0.1cm}}
\newcommand{\spz}{\hspace{0.7cm}}
\newcommand{\virg}{\spz ,\spz}

\newcommand{\th}{\theta}
\newcommand{\al}{\alpha}
\newcommand{\s}{\sigma}

\newcommand{\D}{\Delta}
\newcommand{\goto}{\rightarrow}
\newcommand{\lab}{\label}

\newcommand{\vp}{\varphi}
\newcommand{\ra}{\rangle}
\newcommand{\la}{\langle}
\newcommand{\Ga}{\Gamma}
\newcommand{\ap}{\approx}

\newcommand{\cM}{{\cal M}}

\newfont{\twelvemsb}{msbm10 scaled\magstep1}
\newfont{\eightmsb}{msbm8}
\newfam\msbfam
\textfont\msbfam=\twelvemsb
\scriptfont\msbfam=\eightmsb
\catcode`\@=11
\def\Bbb{\ifmmode\let\next\Bbb@\else
  \def\next{\errmessage{Use \string\Bbb\space only in math mode}}\fi\next}
\def\Bbb@#1{{\fam\msbfam{{#1}}}}

\newcommand{\bC}{{\Bbb C}}

\begin{document}
\setcounter{page}{0}
\topmargin 0pt
\oddsidemargin 5mm
\renewcommand{\thefootnote}{\arabic{footnote}}
\newpage
\setcounter{page}{0}
\begin{titlepage}
\begin{flushright}
ISAS/EP/2000/80
\end{flushright}
\vspace{0.5cm}
\begin{center}
{\large {\bf Universal Amplitude Ratios of 
The Renormalization Group: \\
Two--Dimensional Tricritical Ising Model}} \\
\vspace{1.8cm}
{\large D. Fioravanti$^{a,b}$, G. Mussardo$^{b,c}$ and P. Simon$^{a,b}$} \\
\vspace{0.5cm}
{\em $^{a}$International School for Advanced Studies, Trieste, Italy}\\
\vspace{0.3cm}
{\em $^{b}$Istituto Nazionale di Fisica Nucleare, Sezione di Trieste}\\
\vspace{0.3cm}
{\em $^{c}$Dipartimento di Fisica, Universit\`a dell'Insubria, Como, Italy}
\end{center}
\vspace{1.2cm}

\renewcommand{\thefootnote}{\arabic{footnote}}
\setcounter{footnote}{0}

\begin{abstract}
\noindent
The scaling form of the free--energy near a critical point 
allows for the definition of various thermodynamical amplitudes and 
the determination of their dependence on the microscopic non--universal 
scales. Universal quantities can be obtained by considering 
special combinations of the amplitudes. Together with the critical 
exponents they characterize the universality classes and may be 
useful quantities for their experimental identification. We compute 
the universal amplitude ratios for the Tricritical Ising Model in 
two dimensions by using several theoretical methods from  
Perturbed Conformal Field Theory and Scattering Integrable Quantum 
Field Theory. The theoretical approaches are further supported and 
integrated by results coming from a numerical determination of the 
energy eigenvalues and eigenvectors of the off--critical systems in 
an infinite cylinder. 
\end{abstract}

\vspace{.5cm}

\hspace{5mm} PACS numbers: 64.60.Fr, 05.50+q,75.10Hk
\end{titlepage}

\newpage
\noindent
\resection{Introduction}

One of the most powerful and fascinating concepts in the 
investigation of critical phenomena -- which has successfully 
passed the scrutiny of both experimental and theoretical tests 
during the last decades -- goes under the name of {\em universality} 
\cite{universality}. According to this principle two statistical 
models, which share the same symmetry of the order parameters 
and the dimensionality of the space of their definition, show 
an identical critical behavior although they may greatly differ 
in their microscopic realizations: near the phase transition, 
when the correlation length is much larger than any other 
microscopic scale, they appear as two representatives 
of the same universality class. The first characteristic of 
a given universality class consists of a set of critical 
exponents. Their values are generally given in terms of 
algebraic expressions of the anomalous dimensions of the 
relevant operators present at the critical point. In two 
dimensions, by the powerful methods of Conformal Field Theory 
(CFT) \cite{BPZ,FQS1,DiFMS,DF} one can ascertain both the values 
of the critical exponents, the Operator Product Expansions 
(OPE) and the multi--point correlators of the relative fields. 

However, a complete analysis of the class of universality 
should also include the description of the structure of 
the Renormalization Group trajectories near the critical 
point. The most ambitious goal would be the determination of 
both the scaling function which fixes the equation of state 
and the off--critical correlators of the various order 
parameters\footnote{Near the critical point all quantities 
relative to different models of the same class of 
universality become identical provided an opportune 
rescaling of the order parameters, the external fields and 
the correlation length of the models is made.}. Although 
the exact determination of the equation of state of a 
given universality class may often be a difficult task, 
the scaling property alone of the free--energy is nevertheless 
sufficient to extract numerous predictions on universal 
combinations of critical amplitudes. As it will become clear 
in section 3, these universal combinations are pure 
numbers which can be extremely useful for the identification 
of the universality classes. In fact, the amplitude ratios 
are numbers which typically present significant variations 
between different classes of universality, whereas the critical 
exponents usually assume small values which only vary by a small 
percent by changing the universality classes. Hence the universal 
ratios may be ideal experimental marks of the critical scaling 
regime \cite{Privman,SFW}.  

In recent years, due to the theoretical progress achieved in the 
study of two--dimensional models, some universal ratios and other 
universal quantities have been computed for a large variety of 
bidimensional systems, such as the self--avoiding walks 
\cite{CMpol}, the Ising model [9--14] 
or the q--state Potts 
model \cite{JLC,DelfinoCardy,Caselle2}, to name few. In this paper 
we will concentrate our attention on the determination of the 
amplest possible set of universal amplitude ratios relative to 
the class of universality of the two--dimensional Tricritical 
Ising Model (TIM). Preliminary results relative to some off--critical 
phases of this class of universality have been presented in our 
previous publication \cite{prl} and the aim of this paper is twofold. 
First of all, to complete the list of universal amplitude ratios of the 
TIM presented in \cite{prl} and to perform an exhaustive analysis of 
all its possible phases. Secondly, to illustrate in full detail 
the theoretical methods which have been employed in such determination: 
in view of their successful applications, these techniques may be 
useful to analyze and to obtain similar results for other statistical 
models, in such a way to bridge a closer contact between theoretical 
and experimental results in two--dimensional physics. 

Introducing the Tricritical Ising Model, from a field 
theoretical point of view this model may be regarded as a 
Landau--Ginzburg (LG) $\Phi^6$--theory near its tricritical 
point \cite{ZamLG}. The LG terminology allows a qualitative 
understanding of the phase--structure of the model effortlessly. 
However the LG approach is often too elementary for the
understanding of some remarkable symmetries present in 
the 2-D TIM. As a matter of fact, the bidimensional TIM 
is an unique example of critical phenomena: it is still 
sufficiently simple to be solved but at the same time it 
presents an extremely rich and fascinating structure of 
excitations which can attract the curiosity of a theorist. 
Depending on the direction in the phase space in which the 
system is moved away from criticality, one can observe, 
for instance, a behavior ruled by the exceptional root system 
$E_7$ \cite{EY,MC,FZ} or by supersymmetry 
\cite{Kastor,Zammassless,DMSmassless,Zamthree} (in its exact 
or broken phase realization) or by an asymmetrical pair of 
kinks \cite{LMC,Smirnov12,CKM}. In addition, the description 
of its low--temperature phase is easily obtained from the one of 
its high--temperature phase because of the self--duality of 
the model. From the experimental point of view, a number of physical 
systems exhibit a tricritical Ising behavior, among them fluid mixtures 
or metamagnets\footnote{The interest reader may find ample review 
of this topics in ref.\,\cite{Lawrie}.}. Hence, there is an obvious 
interest in computing the ampler set of data for this class of 
universality and in testing the theoretical predictions 
versus their experimental determination. Our calculation of the 
universal ratios of the TIM will be performed by a combined use of 
results coming from Perturbed Conformal Field Theory, from the 
integrable structure of some of the deformations of the critical 
point action and also from some numerical approaches. 

The paper is organized as follows. In Section 2 we briefly 
describe the universality class of the TIM and the symmetry 
properties of the theories resulting from the deformations of 
the critical point action made of each of the four relevant 
fields. Section 3 is devoted to the discussion of the scaling 
behavior of the singular part of the free--energy and the 
definition of the universal ratios obtained by considering 
some particular combinations of the thermodynamical amplitudes. 
In Section 4 we discuss the Quantum Field Theory approach to 
the computation of the universal ratios. The numerical method 
based on the diagonalization of the off--critical hamiltonian 
obtained by truncating the Hilbert space of the conformal 
space is discussed in Section 5. A thorough analysis of 
each relevant perturbation of the TIM is performed in 
Section 6. Finally our conclusions are presented in 
Section 7. Several appendices, relative to some technical 
aspects of our calculations, are included at the end of 
the paper.  

\resection{The Class of Universality of the Tricritical Ising Model}

In this section we will briefly outline the main properties of 
the universality class of the two--dimensional Tricritical 
Ising model, whereby the detailed discussion of its physical 
properties relative to each of its perturbations is contained 
in Section 6. 

A possible lattice realization of the Tricritical Ising model 
is provided by the so--called Blume--Capel model \cite{Blume}. 
This involves two statistical variables at each lattice site, 
$s_k$ -- the spin variable -- which assumes values $\pm 1$ and 
$t_k$ -- the vacancy variable -- with values $0$ or $1$, which 
specifies therefore if the site is empty or occupied. It is 
characterized by the most general Hamiltonian with nearest 
neighbor pair interaction  
\EQ
{\cal H} = -J \sum_{\langle i,j\rangle}^N 
s_i s_j t_i t_j + \Delta \sum_{i=1}^N t_i - 
H \sum_{i=1}^N s_i t_i -H_3 \sum_{\langle i,j\rangle}^N 
(s_i t_i t_j + s_j t_j t_i) - K \sum_{\langle i,j\rangle}^N 
t_i t_j \,\,\, .
\label{BlumeCapel}
\EN 
The parameter $H$ represents an external magnetic field, 
$H_3$ an additional subleading magnetic source, $J$ the 
coupling between two nearest occupied sites, $\Delta$ 
the chemical potential coupled to the vacancies and $K$ 
an additional subleading energy term between them. Another 
possible two--dimensional lattice realization of this class of 
universality is provided by the so--called dilute $A_L$ models, 
discussed in \cite{dilute}. 

By adopting a field theoretical point of view, a convenient 
way to analyze the universality behavior of the TIM consists 
in considering a Landau--Ginzburg formulation based on a scalar 
field $\Phi(x)$ \cite{ZamLG}. The main advantage of this approach 
is an account of the $Z_2$ symmetry properties of each order 
parameter which provides an easy way of understanding the phase 
structure of the model, at least qualitatively. In this formulation, 
the class of universality of the TIM is associated to the Euclidean 
action 
\EQ
{\cal A} = \int d^Dx \left[\frac{1}{2} (\partial_{\mu} \Phi)^2 + 
g_1 \Phi + g_2 \Phi^2 + g_3 \Phi^3 + g_4 \Phi^4 + 
\Phi^6 \right]\,\,\, ,
\label{LG}
\EN 
with the tricritical point identified by the bare conditions 
$g_1=g_2=g_3=g_4=0$. In a close comparison with the Blume--Capel 
lattice formulation of the model, the statistical interpretation of 
the coupling constants is as follows: $g_1$ plays the role of 
an external magnetic field $h$, $g_2$ measures the displacement 
of the temperature from its critical value $(T-T_c)$, $g_3$ 
may be regarded as a staggered magnetic field $h'$ and finally 
$g_4$ may be thought as a chemical potential for the vacancy 
density. 

A dimensional analysis shows that the upper critical dimension of the 
above LG model is $D=3$, where the tricritical exponents are expected 
to take their classical values, excluding logarithmic corrections. The mean 
field solution of the model easily shows that the LG action (\ref{LG}) 
has a tricritical point, {\it i.e.} a critical point where a line of a 
second order phase transition meets a line of a first order phase transition. 
Consider in fact the case where all the $Z_2$ odd couplings are equally 
set to zero. The potential in this subspace is given by 
\EQ
V(\Phi) = g_2 \Phi^2 + g_4 \Phi^4 + \Phi^6 \,\,\, .
\label{evenLG}
\EN 
The line of a second order phase transition is identified by 
the condition 
\EQ
g_2 = 0 \,\,\,\,\, ,
\,\,\,\,\, g_4 > 0 \,\,\, , 
\label{second}
\EN 
whereas the line of a first order phase transition (where 
three degenerate vacua coexist) is obtained by the condition 
\EQ
g_2 > 0 \,\,\,\,,\,\,\,\,\,
g_4 = -2 \sqrt{g_2} \,\,\,.
\label{first}
\EN 
Hence the point $g_1= g_2 = g_3 = g_4 = 0$ is indeed a tricritical 
point. 

In two dimensions -- the case which will mostly concern us -- there 
are strong fluctuations of the order parameters and therefore the 
exponents and the amplitudes extracted by its mean field solution 
cannot be trusted. However, in this case one can take advantage 
of the powerful methods of Conformal Field Theory to obtain 
an exact solution of this model at criticality. In fact, the 
bidimensional TIM is described by the second representative 
of the unitary series of minimal models of CFT \cite{BPZ,FQS1,DiFMS}: 
its central charge is equal to $c = \frac{7}{10}$ and the Kac-table 
of the exact conformal weights of the scaling fields 
\EQ
\Delta_{l,k} = \frac{(5 l - 4 k)^2 -1}{80} \,\,\,\,\,\,\,,
\,\,\,\,\,
\begin{array}{c}
1 \leq l \leq 3 \\
1 \leq k \leq 4 
\end{array}
\label{Kactable}
\EN
is given in Table 1. There are six primary scalar fields 
$\phi_{\Delta,\overline\Delta}$, which close an algebra under 
the Operator Product Expansion
\EQ
\phi_i(z_1,\overline z_1) \,
\phi_j(z_2,\overline z_2) \,
\sim \,
\sum_k \, c_{ijk} \mid z_1 - z_2 
\mid^{-2 (\Delta_i + \Delta_j - \Delta_k)} \phi_k(z_2,\overline z_2) 
\,\,\,.
\EN
The skeleton form of this OPE algebra and the relative structure 
constants of the Fusion Rules of the TIM are in Table 2. 
The six primary fields can be identified with the normal ordered 
composite LG fields \cite{ZamLG} (see Table 3). With respect to 
their properties under the $Z_2$ spin--reversal transformation 
$Q: \Phi \rightarrow - \Phi$ we have: 
\begin{enumerate}
\item two odd fields: the leading magnetization operator 
$\sigma = \phi_{\frac{3}{80},\frac{3}{80}} \equiv \Phi$ 
and the subleading magnetization operator $\sigma' = 
\phi_{\frac{7}{16},\frac{7}{16}} \equiv : \Phi^3:$ 
\item four even fields: the identity operator $1 = \phi_{0,0}$, 
the leading energy density $\varepsilon = \phi_{\frac{1}{10},
\frac{1}{10}} \equiv :\Phi^2:$, the subleading energy density 
$t = \phi_{\frac{6}{10},\frac{6}{10}} \equiv :\Phi^4: $, 
which in metamagnets assumes the meaning of the density of the 
annealed vacancies, and the field $\varepsilon" = \phi_{\frac{3}{2},
\frac{3}{2}}$. The OPE of the even fields form a subalgebra of 
the Fusion Rules. 
\end{enumerate}
In the TIM there is another $Z_2$ transformation -- the 
Kramers--Wannier duality $ D$ -- under which the fields 
transform as follows: 
\begin{itemize} 
\item 
the order magnetization operators are mapped onto their 
corresponding disorder operators 
\EQ
\mu =  D^{-1} \sigma  D = \tilde\phi_{\frac{3}{80},
\frac{3}{80}} 
\,\,\,\,\, , 
\,\,\,\,\,  
\mu' =  D^{-1} \sigma' D =
\tilde\phi_{\frac{7}{16},\frac{7}{16}} 
\,\,\,.
\label{sigmadual}
\EN 
\item 
the even fields are mapped onto themselves, 
\EQ
D^{-1} \varepsilon D =-\varepsilon \hspace{3mm},
\hspace{3mm} 
D^{-1} t D = t  \hspace{3mm}, \hspace{3mm} 
D^{-1} \varepsilon" D = - \varepsilon" , 
\EN
{\it i.e.} $\varepsilon$ and $\varepsilon"$ are odd under this 
transformation whereas $t$ is even. 
\end{itemize}

Interestingly enough, at criticality the TIM also provides a 
concrete realization of a supersymmetric theory since it is 
the first representative of the superconformal minimal models 
\cite{FQS2,Qiu,MSS}: the even fields can be grouped into a 
superfield of the Neveu--Schwartz sector
\EQ
{\cal N}(z,\bar z,\theta,\bar {\theta})=
\varepsilon(z,\bar z)+
\bar {\theta}\hspace{1mm}
\psi(z,\bar z)+
\theta\hspace{1mm}\bar{\psi}(z,\bar z)+
\theta\bar {\theta}\hspace{1mm}
t(z,\bar z)\hs , 
\label{superfield}
\EN
(where $\theta$ and $\bar{\theta}$ are Grassman variables) while the 
magnetic fields give rise to two irreducible representations 
in the Ramond sector\footnote{The disorder fields $\mu$, $\mu'$ and 
the fermionic fields $\psi$, $\overline\psi$ enter the partition 
function of the model on a torus with twisted boundary conditions, 
see for instance \cite{LMC} where the relative Fusion Rules are 
also presented.}. The critical superconformal LG action is 
given by 
\EQ
{\cal A} = \int d^2x \, d^2\theta \,
\left[ \frac{1}{2} {\cal D} {\cal N} \,
{\bar {\cal D}} {\cal N} 
+ {\cal N}^3 \right]\,\,\, , 
\label{super}
\EN 
with the covariant derivatives defined as 
\EQ
{\cal D} = \frac{\partial}{\partial \theta} - \theta 
\frac{\partial}{\partial z} 
\,\,\,\,\, ,
\,\,\,\,\, 
\overline{{\cal D}} = \frac{\partial}{\partial \overline{\theta}} - 
\overline{\theta} \frac{\partial}{\partial \overline{z}} 
\,\,\,.
\EN

\vspace{3mm}

At the critical point, the TIM can be also realized in terms of 
a coset construction of a Wess-Zumino-Witten model on the group 
$G/H$ given by $(E_7)_1 \otimes (E_7)_1/(E_7)_2$ (for a general 
discussion on the coset model see, for instance \cite{DiFMS}). 
For the central charge $c = c_G - c_H$ we have in fact 
$c = 2 \times 133 \left(\frac{1}{19} - \frac{1}{20}\right) = 
\frac{7}{10}$. Concerning the irreducible representations, for 
the WZW based on $(E_7)_1$ we have $\{{\bf I},{\bf \Pi}_6\}_1$ 
with conformal dimensions $\{0,\frac{3}{4}\}$, whereas for 
the WZW model $(E_7)_2$ we have $\{{\bf I},{\bf \Pi}_1,
{\bf \Pi}_5,{\bf \Pi}_6,{\bf \Pi}_2\}_2$ with conformal dimensions
$\{0,\frac{9}{10},\frac{7}{5},\frac{57}{80},\frac{21}{16}\}$. 
Therefore, the conformal fields of the TIM emerge from the 
decomposition 
\begin{eqnarray}
&& ({\bf I})_1 \times ({\bf I})_1  =  
(I)_{TIM} \otimes ({\bf I})_2 + 
\left(\frac{1}{10}\right)_{TIM} \otimes 
({\bf \Pi}_1)_2 + \left(\frac{6}{10}\right)_{TIM} \otimes 
({\bf \Pi}_5)_2 \nonumber \\
&& ({\bf I})_1 \times ({\bf \Pi}_6)_1  = 
\left(\frac{7}{16}\right)_{TIM} \otimes 
({\bf \Pi}_7)_2 + \left(\frac{3}{80}\right)_{TIM} \otimes 
({\bf \Pi}_6)_2 \label{E7WZWM} \\
&& ({\bf \Pi}_6)_1 \times ({\bf \Pi}_6)_1  =  
\left(\frac{3}{2}\right)_{TIM} \times ({\bf I})_2 \nonumber 
\end{eqnarray}

As will be discussed later, the above symmetries present at 
the critical point of the TIM are also useful for the 
investigation of some off--critical phases of the model. 
The four fields $\sigma$, $\varepsilon$, $\sigma'$ and $t$ 
of increasing anomalous dimensions are, from a Renormalization 
Group point of view, all relevant operators (i.e. their 
conformal weight satisfies $\Delta < 1$) and therefore 
they can be used to move the TIM away from criticality. 
To simplify the formulae below, it is convenient 
to adopt the compact notation $\varphi_i$ $(i=1,2,3,4)$ to 
denote collectively all these fields, so that $\varphi_1 
= \sigma$ , $\varphi_2 = \varepsilon$, $\varphi_3=\sigma'$ 
and $\varphi_4=t$. In the vicinity of the critical point 
the partition function of the model can be expressed by 
the path integral 
\EQ
Z[g_1,g_2,g_3,g_4] = 
\int {\cal D}\phi \,e^{- \left[{\cal A}_{CFT} + 
\sum_{i=1}^4 g_i \int \varphi_i(x)\, d^2x \right]} 
\, \equiv e^{-\hat f(g_1,g_2,g_3,g_4)} \,\,\,.
\label{partitionfunction}
\EN 
We use the notation $Z_1[g_1] = Z[g_1,0,0,0]$, 
$Z_2[g_2] = Z[0,g_2,0,0]$, etc. for the partition functions
corresponding to the individual deformations of the conformal 
action. An immediate result for the off--critical phases can 
be drawn from the symmetry properties of the fields $\varphi_i$. 
In fact, since the fields $\varphi_1$ and $\varphi_3$ are odd 
under the spin--reversal $Z_2$ symmetry, a change of the sign 
of the corresponding coupling constants gives rise to identical 
physical situations, {\it i.e.} $Z_1[g_1] = Z_1[-g_1]$ and 
$Z_3[g_3] = Z_3[-g_3]$. The operator $\phi_2$ is odd under 
the $Z_2$ duality transformation and therefore the physical 
situations which originate from a change of sign of this 
coupling constant will be related by a duality transformation. 
Finally the operator $\phi_4$ is even under both $Z_2$ symmetries  
and therefore the changing of the sign of the corresponding 
coupling constant will produce two distinct physical situations. 

At this stage it is also useful to anticipate the nature of 
Quantum Field Theories which originate from each individual 
deformation, postponing their detailed discussion in Section 6.  
The QFT associated to the deformation of the field $\varphi_1$ 
alone is not integrable: numerical indications of this fact 
were discussed in \cite{LMC}. The QFT which originates from 
the deformation of $\varphi_2$ is instead integrable and the 
pattern of its bound states and the scattering amplitudes are 
related to the hidden $E_7$ algebraic structure of the model 
\cite{MC,FZ}. The deformation of the critical action by means of 
the field $\varphi_3$ produces an integrable field theory made of 
kinks, which interpolate between two asymmetric vacua 
\cite{Smirnov12,CKM}. Finally, the $\varphi_4$ deformation made 
with a positive value of the relative coupling constant corresponds 
to an integrable massless RG flow between the TIM and the standard  
Ising model \cite{Zammassless,DMSmassless} whereas the $\varphi_4$ 
deformation with a negative value of $g_4$ gives rise to an integrable 
massive QFT with kink excitations interpolating between three 
degenerate vacua \cite{Zamthree}. Both these last QFT's give 
explicit realizations of a supersymmetric system in its broken 
and unbroken phase respectively. A useful summary of the 
theories resulting from each deformation can be found in Table 4. 
In conclusion, excluding the magnetic $\varphi_1$ deformation, all the 
others gives rise to integrable QFT. This fact will be quite 
important for our future considerations. 

\resection{Scaling Form of the Free--Energy and Universal Ratios}

The scaling property of the relevant fields which span the 
scaling region of the TIM near its critical point allows  
the derivation of a large set of universal quantities which 
are of experimental interest. For their derivation we will consider 
the general case of the Tricritical Ising Model defined in a 
$D$ dimensional space, even though our final attention will be 
focalized on the two--dimensional system. At this stage of the 
discussion we do not take into account the eventual logarithmic 
corrections explicitly present in the $3$--dimensional version of 
this model\footnote{On this issue the interested reader may 
consult for instance \cite{Lawrie} and references therein.}
and the possible ultraviolet renormalization effects. These, 
however, will be considered within the context of Sections 4 
and 6.  

The scaling property of the order parameters is encoded into 
the asymptotic form of their two--point functions\footnote{This 
notation is not standard for a generic $D$ dimensional system but 
it has the advantage of an easy comparison with formulae which are 
valid in 2-D systems.}
\EQ
\langle \varphi_i (x) \varphi_i(0) \rangle 
\simeq \frac{{\cal A}_i}{\hspace{1mm}|x|^{4 \Delta_i}} \,\,\,\,\,,
\,\,\,\,\, |x|\rightarrow 0 
\,\,\,, 
\label{shortdistance}
\EN 
which therefore identifies the parameters $\Delta_i$ as the conformal 
dimensions of the fields. The standard conformal normalization 
of the fields is obtained by the choice ${\cal A}_i =1$. From the 
power law behavior of (\ref{shortdistance}) it follows 
that the coupling constants $g_i$ behave as 
\EQ
g_i \sim \Lambda ^{D - 2 \Delta_i} \,\,\, ,
\label{coupling}
\EN 
where $\Lambda$ is a mass scale. Therefore, moving the system
away from criticality by means of one of the relevant field 
$\varphi_i$, there will be generally a finite correlation 
length $\xi$ which in the thermodynamical limit scales as 
\EQ
\xi \sim a \,(K_i g_i)^{-\frac{1}{D - 2 \Delta_i}} \,\,\,, 
\label{xi}
\EN 
where $a \sim \Lambda^{-1}$ may be regarded as a microscopic 
length scale. The terms $K_i$ are metric, non--universal factors 
which depend on the unit chosen for measuring the external sources 
$g_i$, alias on the particular realization selected for representing 
the universality class. In the presence of several deformations of 
the conformal action, the most general expression for the scaling 
form of the correlation length may be written as 
\EQ
\xi =\xi_i \equiv a \,(K_i g_i)^
{-\frac{1}{D -2 \Delta_i}}\, 
{\cal L}_i\left(\frac{K_j g_j}{(K_i g_i)^{\phi_{ji}}}\right) \,\,\,, 
\label{xii}
\EN 
where 
\EQ
\phi_{ji} \equiv \frac{D-2 \Delta_j}{D-2 \Delta_i} 
\label{phiji}
\EN 
are the so--called {\em crossover exponents} whereas ${\cal L}_i$ 
are universal homogeneous scaling functions of the ratios 
$\frac{K_j g_j}{(K_i g_i)^{\phi_{ji}}}$. There are of course 
several (but equivalent) ways of writing these scaling forms, 
depending on which coupling constant is selected as a prefactor.  
In the limit where $g_l \rightarrow 0$ ($l \neq i$) but 
$g_i \neq 0$, equation (\ref{xii}) becomes  
\EQ
\xi_i = a \,\xi_i^0 \,g_i
^{-\frac{1}{D -2 \Delta_i}} 
\,\,\,\,\,\,\,\,,
\,\,\,\,\,\,\,
\xi_i^0 \sim K_i^{-\frac{1}{D -2 \Delta_i}} \,\,\,.
\label{xio}
\EN 

Consider now the free--energy $\hat f[g_1,\ldots,g_4]$. This is 
a dimensionless quantity defined by 
\EQ
Z[g_1,g_2,g_3,g_4] = 
\int {\cal D}\phi \, e^{- \left[{\cal A}_{CFT} + 
\sum_{i=1}^4 g_i \int \varphi_i(x) d^2x \right]} 
\, \equiv e^{-\hat f(g_1,g_2,g_3,g_4)} \,\,\,.
\label{free}
\EN 
Assuming the validity of the hyperscaling hypothesis, in the 
thermodynamical limit its singular part (per unit of volume) 
will be proportional to the $D$ power of the correlation length. 
Let us denote the singular part of the free--energy for unit 
volume by $f[g_1,\ldots,g_4]$. Depending on which scaling form 
is adopted for the correlation length, we have correspondingly 
several (but equivalent) ways of parameterizing this quantity 
\EQ
f[g_1,\ldots,g_4] = f_i[g_1,\ldots,g_4] \equiv 
\left(K_i g_i\right)^{\frac{D}{D-2\Delta_i}} \,
{\cal F}_i\left(\frac{K_j g_j}{(K_i g_i)^{\phi_{ji}}}\right) 
\,\,\,. 
\label{scalingfree}
\EN 
The functions ${\cal F}_i$ are universal homogeneous scaling 
functions of the ratios $\frac{K_j g_j}{(K_i g_i)^{\phi_{ji}}}$. 
As it will soon become clear, there is an obvious advantage in 
dealing with different but equivalent expressions for the
free--energy: in fact, since we will be mostly concerned with 
the physical situations which originate from pure deformations 
({\it i.e.} those obtained by keeping only one coupling constant 
finally different from zero), the choice of which one has 
to be selected naturally follows from the particular deformation 
which is considered. 

Let us discuss now the definition of the thermodynamical quantities 
related to the various derivatives of the free--energy. We will 
adopt the notation $\langle ... \rangle_i$ to denote expectation 
values computed in the off--critical theory obtained by keeping 
(at the end) only the coupling constant $g_i$ different from zero. 
The first quantities to consider are the vacuum expectation values 
(VEV) of the fields $\varphi_j$ which can be parameterized as 
\EQ
\langle \varphi_j \rangle_i = -\left.\frac{\partial f_i}{\partial g_j} 
\right|_{g_l=0} \equiv B_{ji} 
g_i^{\frac{2 \Delta_j}{D-2 \Delta_i}} \,\,\,,
\label{vacuumji}
\EN 
with 
\EQ
B_{ji} \sim K_j K_i^{\frac{2 \Delta_j}{D-2 \Delta_i}} \,\,\,.
\label{bji}
\EN 
The above relations can be equivalently expressed as 
\EQ
g_i = D_{ij} \left(\langle \varphi_j\rangle_i\right)^
{\frac{D-2 \Delta_i}{2 \Delta_j}} \,\,\,,
\label{magn}
\EN 
with 
\EQ
D_{ij} \sim \frac{1}{K_i K_j^{\frac{D-2 \Delta_i}{2 \Delta_j}}} 
\,\,\,.
\label{dij}
\EN 
The generalized susceptibilities of the model are defined by 
\EQ
\hat \Gamma_{jk}^i = 
\frac{\partial}{\partial g_k} \langle \varphi_j\rangle_i =\left. 
-\frac{\partial^2 f_i}{\partial g_k \partial g_j}\right|_{g_l=0} 
\,\,\,.
\label{susc}
\EN 
They are obviously symmetrical in the two lower indices. By 
extracting their dependence on the coupling constant $g_i$, 
they can be expressed as 
\EQ 
\hat \Gamma_{jk}^i = \Gamma_{jk}^i \,g_i^
{\frac{2 \Delta_j + 2 \Delta_k - D}
{D - 2 \Delta_i}} \,\,\, ,
\label{hatgammajki}
\EN 
with 
\EQ
\Gamma_{jk}^i \sim K_j K_k K_i^{\frac{2 \Delta_j + 2 \Delta_k - D}
{D - 2 \Delta_i}} \,\,\, . 
\label{gammajki}
\EN 
Some of the above quantities have, of course, a very familiar 
meaning. For instance $\langle \varphi_1\rangle_i$ is nothing 
but the mean value of the magnetization in the off--critical 
theory defined by the $i$--th deformation while $\hat \Gamma^{i}_{11}$ 
is the associated magnetic susceptibility. Similarly, $\langle 
\varphi_2\rangle_i$ is the mean value of the energy along the 
$i$--th deformation of the critical theory and $\hat \Gamma^{i}_{22}$ 
the specific heat. 

As easily seen from the above formulae, the various quantities 
obtained by taking the derivatives of the free--energy contain 
metric factors (the quantities $K_i$) which make their values 
not universal. However, it is always possible to consider special 
combinations thereof in such a way that all metric factors cancel 
out. Here we propose the consideration of the following universal 
ratios 
\EQ
(R_c)^i_{jk} = \frac{\Gamma_{ii}^i \Gamma_{jk}^i}{B_{ji} B_{ki}}
\,\,\, ;
\label{Rc}
\EN 
\EQ 
(R_{\chi})^i_j = \Gamma_{jj}^i D_{jj} B_{ji}^
{\frac{D-4 \Delta_j}{2 \Delta_j}} 
\,\,\,;
\label{Rchi}
\EN 
\EQ
R^i_{\xi} = \left(\Gamma_{ii}^i\right)^{1/D} \xi_i^0 \,\,\,;
\label{Rxi}
\EN 
\EQ
(R_A)^i_j = \Gamma_{jj}^i \, D_{ii}^
{\frac{4 \Delta_j + 2 \Delta_i - 2 D}{D-2 \Delta_i}} \, 
B_{ij}^{\frac{2\Delta_j -D}{\Delta_i}} \,\,\, ;
\label{RA}
\EN 
\EQ
(Q_2)^i_{jk} = \frac{\Gamma^i_{jj}}{\Gamma^k_{jj}} 
\left(\frac{\xi_k^0}{\xi_j^0}\right)^{D-4 \Delta_j} \,\,\,.
\label{Q2}
\EN 
These quantities are pure numbers which therefore characterize 
the universality class of the model. Their definitions closely 
follow and generalize the ones relative to the familiar 
Ising model (see, for instance \cite{Privman,DelfinoIsing}). 
Other universal ratios may be defined as well and in fact some 
of them will be considered in Section 6 devoted to the 
analysis of each deformation of the critical point action. 
Since we will individually compute all the important quantities 
involved ($B_{ij}$, $\Gamma^i_{jk}$, etc.), there is really 
no problem in considering other universal combinations, if 
one wishes to do so. It is worth emphasizing that, from an 
experimental point of view, it should be simpler to measure 
universal amplitude ratios rather than critical exponents: in 
fact to determine the former quantities one needs to perform 
several measurements at a single, fixed value of the coupling 
which drives the system away from criticality whereas to determine 
the latter, one needs to make measurements over several decades 
along the axes of the off--critical couplings. Moreover, although 
not all of them are independent, the universal ratios are a 
larger set of numbers than the critical exponents and therefore 
permit a more precise determination of the class of universality. 

\resection{Quantum Field Theory Approach}

Essential quantities of the universal amplitude 
ratios (\ref{Rc}) -- (\ref{Q2}) are the correlation length 
prefactor $\xi_i^0$, the VEV amplitudes $B_{ji}$ and 
the generalized susceptibilities $\Gamma_{jk}^i$. With  
the aim of determining these quantities, in this section 
we will discuss some useful results relative to the 
two--dimensional Quantum Field Theories (QFT) associated 
to the Renormalization Group flows originating by relevant 
deformations of the conformal action. These QFT are, after 
all, particular representatives of the universality class 
of the model and from now on we will only focalise on them 
for the study of the off--critical dynamics. The advantages 
of adopting this approach will soon become evident. 

In the following we assume the fields to be normalized 
according to the conformal normalization, {\it i.e.} 
\EQ
\lim_{x\rightarrow 0} |x|^{4 \Delta_i} \,
\langle \varphi_i (x) \varphi_i(0)\rangle \, = \, 1 \,\,\,, 
\label{CFTnorm}
\EN 
and we denote by ${\cal M}_i^{\pm}$ the QFT associated to 
the action 
\EQ
{\cal A} = {\cal A}_{CFT} \pm g_i \int \varphi_i(x) \,  
d^2 x \,\,\,\, ,\,\,\,\,\, g_i>0 \,\,\,.
\label{perturbedaction}
\EN 
The coupling constant $g_i$ is a dimensional quantity which 
can be related to the lowest mass--gap $m_i = \xi_i^{-1}$ of 
the theory according to the formula 
\EQ
g_i = \tilde{\cal C}_i \,m_i^{2 - 2 \Delta_i} \,\,\, , 
\label{gm}
\EN 
or, equivalently 
\EQ
m_i = {\cal C}_i \,g_i^{\frac{1}{2-2 \Delta_i}} \,\,\,\,\,\,\, , 
\label{mg}
\EN 
with ${\cal C}_i = \tilde{\cal C}_i^{-\frac{1}{2- 2 \Delta_i}}$. 
When the QFT associated to the action (\ref{perturbedaction}) 
is integrable, the pure number ${\cal C}_i$ can be exactly 
determined by means of the Thermodynamical Bethe Ansatz 
\cite{TBA,fateev}. When the theory is not integrable (this is 
the case for the magnetic deformation of the TIM), the constant 
${\cal C}_i$ can be nevertheless determined by a numerical method, 
based on the so--called Truncated Conformal Space Approach 
\cite{YZ}, which will be discussed in Section 5. In conclusion, 
for all individual deformations of the TIM we are able to completely 
set the relationship which links the coupling constant to the mass--gap 
of the theory and therefore we are able to switch freely between 
these two variables. 

Another set of quantities which can be fixed by QFT are the matrix 
elements of the order parameters, the simplest ones being the vacuum 
expectation values. In this case we have 
\EQ
\langle \varphi_j \rangle_i = \tilde B_{ji} \,m_i^{2 \Delta_j} \,\,\,,
\EN 
{\it i.e.} 
\EQ
\langle \varphi_j \rangle_i = B_{ji} \,g_i^{\frac{\Delta_j}
{1- \Delta_i}} 
\,\,\,\,\,\,\,,
\,\,\,\,\,\,\, 
B_{ji} = \tilde B_{ji} \,{\cal C}_i^{2 \Delta_j} \,\,\,.
\label{VEVg}
\EN 
When the theory is integrable, the constant $\tilde B_{ji}$ can 
be fixed exactly, thanks to the results of a remarkable series of 
papers \cite{russian1,russian2}. When it is not integrable, the 
constant $\tilde B_{ji}$ can be nevertheless estimated by means 
of a numerical approach, first proposed in \cite{GM1}, which will 
be reviewed in Section 5. Hence, also in this case, we are able 
to determine completely these quantities. Let us present the 
exact expressions of the VEV for the three integrable deformations 
of the TIM, which are obtained by specializing the formulae of 
ref.\,\cite{russian2}. In the expressions below, the fields are 
labelled by their position $(l,k)$ in the Kac table of the model 
(see eq.\,(\ref{Kactable}) and Table 1) and the lowest mass--gap 
of the different theories is simply denoted by $m$ 
\begin{itemize}
\item For the $\varphi_2$ energy deformation we have 
\EQ
\langle 0_s \mid \Phi_{l,k} \mid 0_s\rangle =  
\frac
{\sin\left(\frac{\pi s}{4} \mid 5 l - 4 k\mid\right)}
{\sin\frac{\pi s}{4}} \,
\left[
\frac{5 \,m \,\pi \Gamma\left(\frac{5}{9}\right)}
{2^{\frac{2}{3}} \sqrt{3} \Gamma\left(\frac{1}{3}\right)
\Gamma\left(\frac{2}{9}\right)} \right]^{2 \Delta_{l,k}} \,
{\cal Q}_{1,2}(5 l - 4 k) \,\,\,,
\label{VEVepsilon}
\EN 
where, for $\mid Re \,\eta \mid < 4$, ${\cal Q}_{1,2}(\eta)$  
is given by the integral 
\begin{eqnarray}
&& {\cal Q}_{1,2}(\eta)  =  \exp\left\{
\int_0^{\infty} \frac{dt}{t} \left(
\frac{\sinh 6 t \,\sinh(t (\eta-1)) \,\sinh(t (\eta+1))}
{\sinh 15 t \,\sinh 10 t \,\sinh 4 t} \times \right. \right. 
\nonumber \\ & & \left.\left.\left(\cosh 18 t + \cosh 8 t - 
\cosh 16 t + \cosh 4 t + 1 \right)  
- \frac{(\eta^2 -1)}{40} e^{-4 t}\right)\right\} 
\,\,\, ,
\label{Qepsilon} 
\end{eqnarray}
and is defined by its analytic continuation outside that domain. 
The index $s$ labels the various vacua and takes different values 
depending on the sign of the coupling constant $g_2$: for $g_2<0$ the 
spin symmetry $\sigma\rightarrow -\sigma$ is spontaneously broken and 
there are two vacua identified by $s =1,3$; for $g_2 > 0$ there is a 
unique ground state, associated to $s = 2$. 
\item For the $\varphi_3$ sub--leading magnetic deformation we have 
\EQ
\langle 0_s \mid \Phi_{l,k} \mid 0_s\rangle =  
\frac
{\sin\left(\frac{\pi s}{5} \mid 5 l - 4 k\mid\right)}
{\sin\frac{\pi s}{5}} \,
\left[
\frac{4 \,m \,\pi \Gamma\left(\frac{8}{9}\right)}
{2^{\frac{2}{3}} \sqrt{3} \Gamma\left(\frac{1}{3}\right)
\Gamma\left(\frac{5}{9}\right)} \right]^{2 \Delta_{l,k}} \,
{\cal Q}_{2,1}(5 l - 4 k) \,\,\,,
\label{VEVsigma'}
\EN 
where, for $\mid Re \,\eta \mid < 5$, ${\cal Q}_{2,1}(\eta)$ 
is given by the integral 
\begin{eqnarray}
&& {\cal Q}_{2,1}(\eta)  =  \exp\left\{
\int_0^{\infty} \frac{dt}{t} \left(
\frac{\sinh 3 t \,\sinh(t (\eta-1)) \,\sinh(t (\eta+1))}
{\sinh 9 t \,\sinh 5 t \,\sinh 8 t} \times \right. \right. \nonumber \\
& & \left.\left.\left(\cosh 9 t + \cosh t - \cosh 11 t + \cosh 5 t +1 \right) -
\frac{(\eta^2 -1)}{40} e^{-4 t}\right)\right\} 
\,\,\, ,
\label{Qsigma'}
\end{eqnarray}
and is defined by its analytic continuation outside that domain. For 
this deformation there are two vacua associated to the values $s = 2, 
4$. 
\item For the $\varphi_4$ vacancy density deformation, in its 
massive phase we have 
\EQ
\langle 0_s \mid \Phi_{l,k} \mid 0_s\rangle =  
\frac
{\sin\left(\frac{\pi s}{5} \mid 5 l - 4 k\mid\right)}
{\sin\frac{\pi s}{5}} \,
\left[
\frac{m \,\sqrt{\pi} \Gamma\left(\frac{7}{2}\right)}
{2} \right]^{2 \Delta_{l,k}} \,
{\cal Q}_{1,3}(5 l - 4 k) \,\,\,,
\label{VEVepsilon'}
\EN 
where, for $\mid Re \,\eta \mid < 4$, ${\cal Q}_{1,3}(\eta)$ 
is given by the integral 
\EQ
{\cal Q}_{1,3}(\eta)  =  \exp\left\{
\int_0^{\infty} \frac{dt}{t} \left(
\frac{\cosh 2 t \,\sinh(t (\eta-1)) \,\sinh(t (\eta+1))}
{2 \cosh t \,\sinh 4 t \,\sinh 5 t} - \frac{(\eta^2 -1)}{40} 
e^{-4 t}\right)\right\} \nonumber 
\label{Qepsilon'}
\EN
and is defined by its analytic continuation outside that domain. 
For the massive phase of this deformation we have three vacua labelled 
by $s = 1, 2, 3$. 
\end{itemize}

As discussed in Section 5, in addition to the above vacuum 
expectation values, a generalization of the numerical 
approach of ref.\,\cite{GM1} often leads to a reasonable 
estimation of the matrix elements of the order parameters between 
the vacuum states and some of the excited states, as for instance 
$
\langle 0 | \varphi_j | A_k \rangle_i 
$ 
where $A_k$ is a one--particle state of mass $M_k$. These 
quantities will be useful for obtaining sensible approximation 
of the large--distance behavior of several correlators. 

Another useful piece of information on the off--critical dynamics 
can be obtained by exploiting the properties of the stress--energy 
tensor $T_{\mu \nu}(x)$. In the presence of the perturbing field 
$\varphi_i$, the trace of the stress--energy tensor is different 
from zero and can be expressed as 
\EQ
\Theta(x) = 2 \pi g_i (2 - 2 \Delta_i) \,\varphi_i \,\,\,.
\label{trace}
\EN 
The vacuum expectation value of $\Theta(x)$ is given by 
\EQ
\langle \Theta \rangle = \tilde w_i \,m_i^2 = w_i\, 
g_i^{\frac{1}{1- \Delta_i}} 
\,\,\,,
\label{vacuumtheta}
\EN
with $w_i =\tilde w_i \,{\cal C}_i^2$. As before, when the theory is 
integrable the constant $\tilde w_i$ can be determined exactly, 
otherwise it can be computed numerically\footnote{Obviously this 
number can be also obtained in terms of the VEV of the field 
$\varphi_i$ and the expression of $\Theta$ given in eq.\,(\ref{trace}).}. 
The trace of the stress--energy tensor enters two useful sum rules 
which link conformal data of the ultraviolet fixed point to 
off--critical quantities. The first of them -- called the $c$--theorem 
sum rule \cite{Zamcth} -- relates the central charge $c$ of the 
ultraviolet theory to the second moment of the two-point 
connected correlation function of $\Theta$ 
\EQ
c = \frac{3}{4\pi} \int d^2x \,|x|^2 \langle \Theta(x)\,
\Theta(0) \rangle_c 
\,\,\, .
\label{ctheorem}
\EN 
For all relevant deformations, it is easy to check that the above 
integral is always convergent. The second sum rule -- called the 
$\Delta$--theorem sum rule \cite{DSC} -- reads 
\EQ
\Delta_j = -\frac{1}{4 \pi \langle \varphi_j \rangle_i} 
\, \int d^2x \,\langle \Theta(x) \,\varphi_j(0) \rangle_i^c 
\,\,\,, 
\label{Deltath}
\EN 
{\it i.e.} it relates the conformal dimension $\Delta_j$ of the 
field $\varphi_j$ to its VEV and to the integral of its connected 
off--critical correlator with $\Theta(x)$. This time the above 
integral is not always convergent (the detailed analysis of 
its convergence may be found in the original paper \cite{DSC}). 
Notice, however, that the $\Delta$--theorem also involves the 
VEV of the field $\varphi_j$ and it is easy to see that the 
divergence/convergence of the integral is always accompanied 
by the divergence/convergence of the VEV in such a way that 
the sum--rule always mantains its validity. The proof of this 
statement is simple: by taking the derivative of the VEV 
(\ref{VEVg}) with respect to $g_i$ we have in fact 
\EQ
\frac{\partial}{\partial g_i} \langle \varphi_j\rangle_i = 
\frac{\Delta_j}{g_i (1-\Delta_i)} \,\langle \varphi_j\rangle_i 
\,\,\,.
\label{derVEV}
\EN 
On the other hand, the above quantity can also be computed by 
means of the fluctuation--dissipation theorem and is given 
by 
\EQ
\frac{\partial}{\partial g_i} \langle \varphi_j\rangle_i = 
- \int d^2 x \langle \varphi_i(x) \varphi_j(0) \rangle_c^i \,\,\,.
\label{derflu}
\EN 
By using eq.\,(\ref{trace}) and comparing the two expressions, 
the divergent/convergent nature of the integral is therefore 
directly linked to the divergent/convergent nature of the 
constant $B_{ji}$ entering the VEV of the field $\varphi_j$. 
These considerations suggest that the quantity $\Delta_j$ 
on the left hand side of (\ref{Deltath}) is in any case 
obtained, also when the integral of the two--point function 
$\langle \Theta(x) \varphi_j(0)\rangle_c^i$ diverges. In this 
case one needs to perform a simultaneous analytic continuation 
of both the integral and the corresponding VEV. 

Basic quantities in the universal ratios are 
the generalized susceptibilities $\Gamma_{jk}^i$. By 
using equations.\,(\ref{free}) and (\ref{hatgammajki}), the 
fluctuation--dissipation theorem provides the relation 
between $\hat\Gamma_{jk}^i$ and the integral of the 
connected correlator 
\EQ
\hat\Gamma_{jk}^i = \int d^2 x \langle \varphi_j(x) 
\varphi_k(0)\rangle_{c}^i 
\,\,\,.
\label{fluctdis}
\EN 
The dependence on the coupling constant $g_i$ of these 
quantities can be easily extracted. In fact, the connected 
correlator can be parameterized as 
\EQ
\langle \varphi_j(x) \varphi_k(0)\rangle_c^i = 
\frac{1}{r^{2 \Delta_j + 2 \Delta_k}} Q_{jk}^i(m \,r) 
\,\,\,.  
\EN 
($r = \mid x \mid$). Its dependence on $m$ is obtained 
with a change of variable and by using the relation
({\ref{mg}), we finally have $\hat \Gamma_{jk}^i = 
\Gamma_{jk}^i\, g_i^{\frac{\Delta_j + \Delta_k-1}{1-\Delta_i}}$ 
with 
\EQ
\Gamma_{jk}^i = {\cal C}_i^{2 \Delta_j+2\Delta_k-2} 
\int d\tau \frac{1}{\tau^{2\Delta_j+2\Delta_k}} Q_{jk}(\tau) 
\,\,\,.
\label{finalgamma}
\EN
Some of the above susceptibilities can be determined exactly, such 
as the components $\Gamma_{ik}^i$, whose values are 
provided by the $\Delta$--theorem sum rule 
\EQ
\Gamma_{ik}^i = - \frac{\Delta_k}{1-\Delta_k} B_{ki} \,\,\,.
\label{gammachiusa}
\EN 
In all other cases, when an exact formula is not available, 
our strategy to evaluate the generalized susceptibilities  
will rely on two different representations of the correlators. 
These representations have the advantage to converge
very fast in two distinct regions: the first representation is 
based on Conformal Perturbation Theory and allows a 
very efficient estimation of the correlation function in its 
short distance regime, while the second representation is based 
on the Form Factors and allows an efficient control of its large 
distance behavior. Due to the fast convergent nature of the two 
series in their respective domains, they are efficiently approximated 
by their lowest terms, which therefore can be evaluated with a 
relatively little analytical effort. These considerations obviously 
lead to the estimation of the integral (\ref{fluctdis}) according 
to the following steps: 
\begin{enumerate}
\item Express the integral in polar coordinates as 
\EQ
\hat\Gamma_{jk}^i = 2\pi \int_0^{+\infty} 
d r \,r \,\langle\varphi_j(r) 
\varphi_k(0)\rangle_c^i \,\,\, , 
\EN 
and split the radial integral into two pieces as 
\begin{eqnarray}
I = \int_0^{+\infty}  
d r \,r\,\langle\varphi_j(r) 
\varphi_k(0)\rangle_c^i & = & 
\int_0^{R}  
d r \,r\,\langle\varphi_j(r) 
\varphi_k(0)\rangle_c^i +  
\int_R^{+\infty}  
d r \,r\,\langle\varphi_j(r) 
\varphi_k(0)\rangle_c^i  \nonumber \\
& & \equiv I_1(R) + I_2(R) \,\,\,.
\label{I1I2}
\end{eqnarray}
\item 
Use the best available short--distance representation of the 
correlator to evaluate $I_1(R)$ as well as the best available 
estimate of its large--distance representation to evaluate 
$I_2(R)$. 
\item Optimize the choice of the parameter $R$ in such a 
way to obtain the best evaluation of the whole integral. 
In practice, this means looking at that value of $R$ 
for which a plateau is obtained for the sum of $I_1(R)$ 
and $I_2(R)$. 
\end{enumerate}
In order to proceed in the above program it is useful to 
briefly recall the main features of the short--distance and 
long--distance expansions of the two--point correlation 
functions.  

\subsection{Short--Distance Expansion}

A clear discussion on the perturbative ultraviolet 
renormalization of the fields and on the short--distance 
expansion of the two--point functions can be found in the 
references \cite{DSC,ZamYL}. Here we will briefly review 
the main results useful for our purposes. 

First of all, there is a one--to--one correspondence between 
the fields {\em at} and {\em away} from criticality. However, 
renormalization effects induced by ultraviolet divergences 
can have the effect of expressing the off--critical fields in 
terms of a combination of the critical ones. Let us denote 
by $\tilde\Phi_i(x)$ and $\Phi_i(x)$ the conformal and the 
off--critical fields respectively. Consider the off--critical action 
obtained by a perturbation of a relevant (scalar) conformal 
field ($\Delta_{\tilde\Phi} < 1$)
\EQ
{\cal A} ={\cal A}_{CFT} + g \int d^2x \,\tilde{\Phi}(x) \,\,\,.
\EN
In a conformal perturbative evaluation of the correlators 
which involves one of the field $\Phi_i(0)$ we have 
\EQ
\langle ... \Phi_i(0)\rangle = 
\langle ... \tilde\Phi_i(0)\rangle_{CFT} + 
g \int_{\epsilon <\mid x\mid < R} 
d^2 x \langle ... \tilde\Phi_i(0) \tilde\Phi(x)\rangle_{CFT} 
+ \ldots 
\label{CFTexpan}
\EN 
where $\epsilon$ and $R$ are the ultraviolet and the infrared 
cutoffs. Let us analyze first the ultraviolet behavior of 
the above integral. This is controlled by the OPE of the 
two conformal fields 
\EQ
\tilde\Phi(x) \tilde\Phi_i(0) = \sum_k C_{\tilde\Phi 
\tilde\varphi}^k \mid x\mid^{2(\Delta_k-\Delta_{\Phi} - 
\Delta_{\Phi_i})} \tilde A_k(0) 
\,\,\,,
\label{CFTOPE}
\EN 
and therefore the integral in (\ref{CFTexpan}) is divergent 
if in the above expansion there are fields $\tilde A_k(x)$ whose 
conformal dimensions satisfy the condition 
\EQ
\gamma_k \equiv \Delta_k - \Delta_{\tilde\Phi} - \Delta_{\tilde\Phi_i} 
+ 1 \leq 0 \,\,\,.
\label{Gescond}
\EN 
If this is the case, the off--critical renormalized field which 
has a finite correlator at the lowest order in $g$ is defined by 
\EQ
\Phi_i = \tilde\Phi_i - g \pi\, \sum_k \frac{C_{\Phi \Phi_i}^k} 
{\gamma_k} \,\epsilon^{2\gamma_k}\,\tilde A_k + {\cal O}(g^2) \,\,\,.
\EN 
Hence, due to the ultraviolet divergences there may be a mixing 
of the initial conformal operators with a finite numbers of fields 
of lower conformal dimensions. 

The conformal perturbation series for the correlation functions 
also suffers from infrared divergences. These divergences, however, 
cannot be absorbed into a redefinition of the local fields and, 
as a result, we have a non--analytic dependence on the coupling 
constant $g$. This non--analytic behavior is essentially due to 
the non--adiabatic change of the vacuum state in passing from 
the Conformal Field Theory of the fixed point to the generic 
massive theory of the off--critical system. To overcome 
this difficulty, one can adopt the strategy of considering the 
{\em off-critical} OPE 
\EQ
\varphi_i(x) \varphi_j(0) = 
\sum_p C_{ij}^p(g;x) A_p(0)\,\,\,,
\label{ope}
\EN
where the $A_p(x)$ belongs to a complete set of local fields 
of the theory ({\it i.e.} the perturbed version of the conformal 
fields $\tilde{A}_p(x)$) whereas the structure constants 
$C_{ij}^p(g;x)$ are analytic in $g$ (as expected by their 
local nature). In this way all the non--analytic behavior 
of the correlation function
\EQ
\langle\varphi_i(x) \varphi_j(0)\rangle = 
\sum_i C_{ij}^p(g;x) \langle A_p(0)\rangle  \,\,\,
\label{correlator}
\EN
is completely encoded inside the non--perturbative VEV's 
$\langle A_p(0)\rangle = {\cal A}_p \,g^{\frac{\Delta_p}
{1-\Delta_{\Phi}}}$. Concerning the structure constants 
$C_{jk}^p(g;x)$, by dimensional reasons they admit the 
expansion
\EQ
C_{ij}^p(g;x)= r^{2(\D_p-\D_i-\D_j)} \,
\sum_{n=0}^{\infty}C_{i,j}^{p(n)}(gr^{2-2\Delta_{\Phi}})^n 
\,\,\, ,
\label{Cexpansion}
\EN
where $r=\mid x\mid$ and they can be computed 
perturbatively\footnote{The above formula has to be 
opportunely corrected when there is a resonance phenomenon 
among the fields of the conformal families.} in $g$. Their 
first order contribution is given by \cite{ZamYL}
\EQ
C_{i,j}^{p(1)} = -\int ' d^2w \,\langle \tilde{A}^p(\infty) 
\tilde{\Phi}(w)\tilde{\varphi}_i(1)\tilde{\varphi}_j(0)\rangle_{CFT} 
\,\,\,,
\label{firstorderc}
\EN
where the prime indicates a suitable infrared (large distance) 
regularization of the integral. It can be calculated by means of 
different approaches (as, for instance, a minimal subtraction 
scheme based on the OPE \cite{ZamYL} or an analytic prolongation 
in the parameters of the integrand of (\ref{firstorderc}), {\it i.e.} 
the conformal weights). As shown in \cite{GM3}, an efficient way to
extract the finite part of the integral is provided by the Mellin
transformation, discussed in Appendix A. This is the approach which 
we have mostly used in our calculations. 

In conclusion, by using the VEV of the fields and the first 
approximation of the structure constants of the OPE, we can 
obtain a reasonable approximation for the short--distance 
behavior of the connected two--point functions entering the 
definition of the susceptibilities in terms of the expression 
\begin{eqnarray}
\langle\varphi_i(x) \varphi_j(0)\rangle^c_k \, &=& 
\sum_p \sum_{n=0}^{\infty} \frac{\langle A_p(0)\rangle_k}
{r^{2 (\Delta_p-\Delta_i-\Delta_j)}} C_{ij}^{p(n)} 
(g r^{2-2 \Delta_{k}})^n   + \nonumber \\ 
& & - \langle \varphi_i\rangle_k \, 
\langle \varphi_j\rangle_k \,\,\,. 
\label{correl}
\end{eqnarray}
The index $k$ indicates which perturbation is considered.
Since the short--distance representation (\ref{correlator}) 
is an expansion in the parameter $\left(\frac{r}{\xi}
\right)$, a truncated form of the series (\ref{correl}) 
is expected to be sufficiently accurate for $r \ll \xi$.  
However, the convergence of the truncated series is often 
much better and results being sufficiently accurate also for 
$r \sim \xi$, as confirmed in several examples (see, for instance 
\cite{DSIMMF,GM3,GM2,ZamYL,DMM35}). Hence, the above truncated 
form (\ref{correl}) can be confidently used for the evaluation 
of the integral $I_1(R)$ in eq.\,(\ref{I1I2}). Finally, notice 
that the short--distance expansion of the correlators can be 
implemented independently on the integrable or non--integrable 
nature of the off--critical theory.

\subsection{Large Distance Expansion}

An efficient way to control the behavior of the correlators 
in the opposite regime $\left(\frac{r}{\xi}\right) \gg 1$ is 
provided by their spectral representation expansions. In this 
approach, one makes use of the knowledge of the off--critical 
mass spectrum of the theory to express the correlators 
as\footnote{The expression (\ref{spectral}) has to be suitably 
modified in presence of kink excitations or massless particles.} 
\EQ
\langle\varphi_i(x) \varphi_j(0)\rangle = \sum_{n=0}^{\infty} 
g_n(r) \,\,\,,
\label{g_n}
\EN 
where 
\begin{eqnarray}
\label{defff}
g_n(r) & = & \int_{\th_1 >\th_2 \ldots>\th_n} 
\frac{d\th_1}{2\pi} \dots \frac{d\th_n}{2\pi}\,
\langle 0|\varphi_i(0)|A_{a_1}(\th_1) \dots A_{a_n}(\th_n)
\rangle  \times \label{spectral} \\
& & \,\,\,\,\,\, \times 
\langle A_{a_1}(\th_1) \dots A_{a_n}(\th_n)|
\varphi_j(0)|0 \rangle \,e^{-r \sum_{k=1}^n m_k \cosh\th_k}
\,\,\,. 
\nonumber
\end{eqnarray} 
$|A_{a_1}(\theta_1) \dots A_{a_n}(\theta_n)\rangle$ are the 
multi--particle states relative to the excitations of mass 
$m_k$, with relativistic dispersion relations given by $E 
= m_k \cosh \theta$, $p = m_k \sinh\theta$, where $\theta$ 
is the rapidity variable. The spectral representation 
(\ref{spectral}) is obviously an expansion in the parameter 
$e^{-\frac{r}{\xi}}$, where $\xi^{-1} = m_1$ is the lowest 
mass--gap. 

Basic quantities of the large distance approach are the Form 
Factors (FF), {\it i.e.} the matrix elements of the operators 
$\varphi_i$ on the physical asymptotic states
\EQ
F^{\varphi_i}_{a_1,\ldots ,a_n}(\th_1,\ldots,\th_n) \,= 
\, \langle 0| \varphi_i(0)|A_{a_1}(\th_1),\ldots,A_{a_n}
(\th_n) \rangle \,\,\,.
\label{form}
\EN
It is worth emphasizing that the above quantities are unaffected 
by renormalization effects since physical excitations are employed 
in their definitions. For scalar operators, relativistic invariance 
requires that the FF only depend on the rapidity differences $\th_i 
- \th_j$. Postponing a more detailed analysis of the analytic 
properties of the FF, let us first discuss the behavior of the 
series (\ref{spectral}) for the purpose of evaluating the integral 
$I_2(R)$ in eq.\,(\ref{I1I2}). 

First of all, it is convenient to order the multi--particle states
entering the sum (\ref{spectral}) according to the increasing values 
of the total sum of their masses $E_n^{a_1,\cdots,a_n} = \sum_{k=1}^n 
m_k$ so that, for $r \gg \xi$, the functions $g_n(r)$ behave 
as the decreasing sequence $g_n(r) \sim e^{-r E_n^{a_1,\cdots,a_n}}$. 
Apart from $g_0 = \langle 0 | \varphi_i(0) | 0 \rangle \langle 0 | 
\varphi_j(0) | 0 \rangle$ relative to the VEV of the fields 
(which however does not enter the connected correlator), the 
first approximation of the correlator is given by the first 
term of the expansion (\ref{spectral}) 
\begin{eqnarray}
\langle \varphi_i(x) \varphi_j(0)\rangle & \simeq & g_1(r) 
= F^{\varphi_i}_1 \, F^{\varphi_j}_1 \, 
\int_{-\infty}^{+\infty} 
\frac{d\theta}{2 \pi} e^{-m_1 r \cosh\theta} = 
\nonumber \\ 
& = & 
F^{\varphi_i}_1 \, F^{\varphi_j}_1 \,\frac{1}{\pi} \, 
K_0(m_1 r) \,\,\, ,
\label{Bessel}
\end{eqnarray} 
where $K_0(x)$ is the modified Bessel function. Sometimes it may 
occur that the one--particle FF of the fields are zero for 
symmetrical reasons and, in this case, the leading approximation of the 
connected correlator is given by the function $g_2(r)$, expressed 
in terms of the two--particle Form Factors $F^{\varphi_i}_{a_1,a_2}
(\theta_1-\theta_2)$. Since (\ref{spectral}) is an exact expansion 
in $e^{-\frac{r}{\xi}}$, its truncated series to the lowest terms 
is expected to provide an accurate approximation of the correlators 
in the interval $r \gg \xi$. However, the convergence property of 
the truncated series is much better \cite{CMpol} and as a matter of 
fact it neatly approximates the correlator up to the region $r \sim \xi$, 
as has been checked in several examples (see, for instance 
\cite{DMIMMF,DSIMMF,GM3,ZamYL,DMM35,AMV}). Therefore the truncated 
spectral series is assumed to estimate the integral $I_2(R)$ 
in eq.\,(\ref{I1I2}) with a reasonable confidence: obviously, the 
more terms included in the series (\ref{spectral}) results in a 
better evaluation of the integral $I_2(R)$. The problem is 
then to determine how efficiently we can assess the matrix elements 
of the order parameters on the asymptotic states. Let us discuss 
separately the cases when the off--critical theory corresponds to a 
non--integrable QFT or to an integrable one. 

For a non--integrable QFT (as, for instance, the QFT resulting from 
the magnetic deformation of the TIM), unfortunately it is difficult 
to go beyond the one--particle Form Factors of the lowest particle 
states, $\langle 0 |\varphi_i(0) | A_k \rangle$. In fact, due to 
creation and annihilation events in the scattering processes of 
these theories, {\it i.e.} to the non--elastic nature of its 
$S$--matrix, the FF satisfy the infinite coupled set of Watson's 
equations \cite{Watson}
\begin{eqnarray}
& & F^{\varphi}_{{\rm in}}(\theta_1),\ldots,\theta_n)  =   
\langle 0 |\varphi(0) | A(\theta_1) \ldots A(\theta_n) \rangle_{{\rm in}} =
\\
& & \sum_{m=0}^{\infty} \int \frac{d\theta_1'}{2\pi} 
\cdots 
\frac{d\theta_m'}{2\pi} 
\langle 0 |\varphi(0) | A(\theta_1') \ldots A(\theta_m') 
\rangle_{{\rm out}} \langle A(\theta_1') \ldots A(\theta_m') 
| A(\theta_1) \ldots A(\theta_n)\rangle_{{\rm in}} = 
\nonumber \\ 
& & \sum_{m=0}^{\infty} \int \frac{d\theta_1'}{2\pi} 
\cdots 
\frac{d\theta_m'}{2\pi} F^{\varphi}_{{\rm out}}(\theta_1',\ldots 
\theta_m') \, S^{n\rightarrow m}(\theta_1,\ldots,\theta_n|
\theta_1',\ldots,\theta_m') \,\,\, , \nonumber 
\end{eqnarray}
obtained by inserting the unitary sum on the out--states in the 
definition of the original Form Factor. Consequently, higher--particle 
FF have a non--trivial analytic behavior -- with branch cuts at all 
production thresholds -- which in practice precludes their exact 
determination. Hence, for non--integrable theories the best we can do 
is to estimate the large--distance expansion of the correlators 
only in terms of the lowest one--particle states. Moreover, 
the one--particle FF of these theories $\langle 0 | \varphi_i(0) 
| A_k\rangle$ cannot be determined by first principles and for 
their evaluation we have to rely on some numerical determinations, 
as discussed in next section. Although this situation may 
appear disappointing from a theoretical point of view, it is 
worth stressing that for all practical purposes, one can reach 
a reasonable estimate of the integral $I_2(R)$ also in the 
non--integrable case. This can be checked, for instance, 
by comparing the values of the integrals -- obtained by approximating 
the correlators by the lowest Form Factors -- with their exact 
values obtained by the $\Delta$--theorem, when the latter applies. 

For an integrable QFT, the situation is much better since, in principle, 
there is the possibility of determining exactly {\em all} Form Factors 
of the theory. For a detailed discussion of the calculation of the 
Form Factors in an integrable QFT, we refer the reader to the original 
literature \cite{DMIMMF,ZamYL,KW,Smirnov}. Here we simply recall  
the basic equations of the two--particle FF \cite{DMIMMF} since, 
based on the fast convergence property expected for the spectral 
series \cite{CMpol}, they will be the only terms employed in the 
following for approximating the correlators in their large--distance 
expansions (in addition, of course, to the one--particle ones). 

Assume, for simplicity, that the spectrum of the integrable QFT is 
made of the scalar particles $A_i$. Let $S_{ab}(\theta)$ be the 
elastic scattering matrix of the particles $A_a$ and $A_b$. In 
this case, the two--particle Form Factor $F^{\varphi}_{ab}(\theta)$  
is a meromorphic function of the rapidity difference $\theta$ 
satisfying the equations 
\EQ
F^{\varphi}_{ab}(\th)=S_{ab}(\th)\,F^{\varphi}_{ab}(-\th)\,\,,
\lab{w1}
\EN
\EQ
F^{\varphi}_{ab}(i\pi+\th) = F^{\varphi}_{ab}(i\pi-\th)\,\,\,.
\lab{w2}
\EN
Let $F^{\it min}_{ab}(\th)$ be a solution of eqs.\,(\ref{w1}) and 
(\ref{w2}), free of poles and zeros in the strip ${\rm Im} \,\theta 
\in (0,\pi)$. By requiring asymptotic power limitation in momenta, 
$F^{\varphi}_{ab}(\th)$ must be equal to $F^{\it min}_{ab}(\th)$ 
times a rational function of $\cosh\th$, with the poles thereof 
fixed by the singularity structure of the scattering amplitude 
$S_{ab}(\th)$. A simple pole in $F^{\varphi}_{ab}(\th)$, like 
the one in Figure 1 induced by the simple pole of 
$S_{ab}(\th)$ with a positive residue, gives rise to the equation 
\EQ
F^{\varphi}_{ab}(\th\simeq
iu_{ab}^c)\simeq\frac{i \gamma_{ab}^c}{\th-iu_{ab}^c} \,
F^{\varphi}_c\,\,\,, 
\lab{pole}
\EN  
where $\gamma_{ab}^c$ is the on--shell three--particle coupling, 
also determined by the $S$--matrix. A more detailed analysis is, in general, 
required when the $S$--matrix presents higher order poles (see the 
discussion in \cite{DMIMMF} and Appendix D). The FF may also have 
kinematical poles which however do not appear at the two-particle 
level if the operator ${\varphi}(x)$ is local with respect to the 
fields which create the particles. In conclusion, 
the two-particle FF can be expressed as \cite{DMIMMF} 
\EQ
F^{\varphi}_{ab}(\th)=\frac{{\cal Q}^{\varphi}_{ab}(\th)}{D_{ab}(\th)}
F^{min}_{ab}(\th)\,\,,
\lab{param}
\EN
where $D_{ab}(\th)$ and ${\cal Q}^{\varphi}_{ab}(\th)$ are 
polynomials in $\cosh\th$: the former is fixed by the singularity 
structure of $S_{ab}(\th)$ while the latter depends on the 
operator ${\varphi}(x)$. An upper bound on the order of the 
polynomial $Q_{ab}^{\varphi}(\th)$ is given in terms of the 
conformal dimension $\Delta_{\varphi}$. In fact  
\EQ
\lim_{|\th|\goto\infty}
F^{\varphi}(\th) \sim \,
e^{y_{\varphi}|\th|}\,\,\,, 
\lab{bound}
\EN
with \cite{DMIMMF} 
\EQ
y_{\varphi}\,\leq\,\Delta_\varphi\,\,\,.
\lab{bbb}
\EN
Further equations on the polynomial ${\cal Q}_{ab}^{\varphi}(\th)$ 
can be obtained when the field $\varphi(x)$ is proportional 
to the trace of the stress--energy tensor $T_{\mu\nu}(x)$. 
In fact, as a consequence of the conservation law $\partial_
\mu T^{\mu\nu}=0$, the FF of $\Theta(x)$ for two different 
particles $A_a$ and $A_b$ must contain a term proportional 
to the Mandelstam variable $s = (p_a+p_b)^2$ of this state 
\cite{DMIMMF,ZamYL}, so that it can be factorized as 
\EQ
{\cal Q}^\Theta_{ab}(\th) = \left(\cosh\th +
\frac{m_a^2+m_b^2}{2m_am_b}\right)^{1-\delta_{ab}} 
P_{ab}(\th)\,\,\,.
\EN
Moreover, in this case we have the normalization conditions 
for $F^\Theta_{aa}$ which reads  
\EQ
F_{aa}^\Theta(i\pi) 
= \langle A_a(\th_a)|\Theta(0)|A_a(\th_a)\rangle = 2\pi m^2_a\,\,\,.
\label{ipi}
\EN

The above discussion relative to the FF (and their generalization 
in the case of kink excitations) will be useful in Section 6 for 
the determination of the generalized susceptibilities for the 
integrable deformations of the TIM. 

\resection{Numerical Methods: Truncated Conformal Space Approach}

The Truncated Conformal Space Approach (TCSA) has been introduced 
by Yurov and Zamolodchikov \cite{YZ} for a numerical evaluation of 
the non--perturbative effects relative to the off--critical models. 
It consists in studying the numerical spectrum of the off-critical 
Hamiltonian on a infinite cylinder of circumference $R$, acting 
on the Hilbert space of the conformal states. Once a truncation 
at a suitable number of states is made, the problem reduces to 
perform a numerical diagonalization of a finite dimensional 
Hamiltonian. 

For the off--critical Hamiltonian we have 
\EQ
\hat{H} = \hat{H}_0 + \hat{V} \,\,\, ,
\label{pert}
\EN
where $\hat{H}_0$ is the Hamiltonian of the conformal fixed point 
on the cylinder and $\hat{V} = g_i \int\limits_0^R dv \,\hat 
{\varphi}_i(w)$ where $\varphi_i(x)$ is one of the relevant 
perturbation (with $w=u+iv$ is the coordinate along the cylinder 
and the tilde indicates quantities defined on the cylinder). By 
using the conformal transformation $z=e^{{2\pi\over R}w}$, the 
conformal theory on the cylinder is mapped onto a plane and 
therefore $\hat{H}_0$ can be expressed in terms of the usual 
conformal generators $L_0,\bar{L_0}$ and the central charge 
$c$ \cite{cardy86}:
\bea
\hat{H}_0 &=& {2\pi\over R}(L_0+\bar{L_0}-{c\over 12}) \,\,\,;\\
\hat{\varphi}_i &=& \left|{2\pi \over R}\right|^{2\D_i}\varphi_i 
\,\,\,.
\eea
The spectrum of $\hat{H}$ depends on the dimensionless parameter
$g_iR^{2-2\D_i}$ and the value of $g_i$ can be fixed\footnote{
By using this normalization, we were able to determine, in particular, 
the constant ${\cal C}_1$ entering the relation between the coupling 
and the mass in the magnetic $\vp_1$ deformation (see eq.\,(\ref{mg})).} 
such that the mass gap is equal to $1$. Let $b^{-1}_{ml}$ be the 
inverse of the matrix $b_{ml} = \langle m | l \rangle$ introduced 
to account for the nonorthogonality of the conformal basis. Denoting 
by $H_{mn} \equiv b^{-1}_{ml} \, \langle l| H |n\rangle$ the matrix 
elements of the perturbed Hamiltonian, we have 
\EQ 
H_{mn}={2 \pi\over R} \left[ (2\Delta_m -c/12) \delta_{mn} +
2 \pi g_i \left({R\over 2 \pi}\right)^{2(1-\Delta_i)} b^{-1}_{ml}
 \langle l|\varphi_i |n\rangle \right] \,\,\,.
\label{htrunc}
\EN
The matrix elements  $\langle l|\varphi_i| n \rangle$ can be computed 
in terms of the structure constants of the OPE and the action of the 
conformal generators $L_n$ on the states. For the perturbation of 
the minimal models of CFT a numerical algorithm has been designed 
to compute the above matrix elements and to perform the diagonalization 
of the off--critical Hamiltonian by including the conformal states 
and their descendants up to the fifth level of the Verma module 
\cite{LM}. In the case of the TIM this is equivalent to truncate 
the number of states $N$ to 228. Once the Hamiltonian $H$ has been 
diagonalized for different values of $R$, one can extract the 
spectrum as a function of $R$, in particular the low energy 
eigenvalues (see, for instance Figure 2) and 
also their associated eigenvectors. In this way, it is possible 
to determine the masses of the lowest particles, several vacuum 
expectation values, some of the one--particle Form Factors and 
also some of the generalized susceptibilities. There are however 
certain limitations of the method, one of them due to the truncation 
performed in the number $N$ of the conformal states employed in 
the algorithm (see the discussion in \cite{LMC}). In fact, all the 
quantities of interest are infrared data, {\it i.e.} relative to 
the dynamics of the system on the cylinder in the limit $R 
\rightarrow \infty$, which is however dominated by truncation 
effects. This means that in order to extract reliable infrared 
data one has to look at the spectrum within an interval of $R$ 
sufficiently large but still unaffected by truncation errors. 
This interval will be called the ``physical window''. Another 
limitation of the TCSA occurs when the conformal dimension 
$\Delta$ of the perturbing operator is such that $\Delta 
\geq \frac{1}{2}$. In this case, in fact, the renormalization of 
the operator prevents to reach a proper scaling behavior of the 
energy levels and the only quantities which can be extracted 
with reasonable confidence are the energy differences  
$\Delta E_n(R) = E_n(R) - E_0(R)$.

\subsection{Vacuum Expectation Values by TCSA}

As shown in ref.\,\cite{GM1}, the knowledge of the eigenvectors 
in the Truncated Conformal Space Approach allows a numerical 
estimation of the Vacuum Expectation Values of several order 
parameters. These quantities are defined by the limit 
\EQ
B_{ji} = \lim_{R\to \infty}\langle \tilde{0}|\hat{\varphi}_j|
\tilde{0} \rangle_i \,|g_i|^{-{\D_j\over 1-\D_i}} \,\,\, ,
\label{IRlimit}
\EN
where $|\tilde{0}\rangle_i$ is the vacuum (on the cylinder) 
relative to the off--critical theory along the $i$--th 
deformation. On the other hand  
\EQ
\label{numvev}
\langle \tilde{0}|\hat{\varphi}_j|\tilde{0}\rangle_i 
= \left({2\pi\over R}\right)^{2\D_j} \langle\tilde{0}|
\varphi_j| \tilde{0} \rangle_i = \left({2\pi\over
R}\right)^{2\D_j}{\psi_m^0
\langle m|\varphi_j| n \rangle_i \psi_n^0\over 
\psi_m^0 b_{mn}  \psi_n^0}\,\,\,,
\EN
where $\psi_m^0$ designs the $m^{th}$ component of the ground state 
vector expressed in terms of the conformal basis. Due to the truncation 
effects discussed above, the limit $R \rightarrow \infty$ in 
eq.\,(\ref{IRlimit}) in practice means that one has to consider 
$R$ large enough such that the VEV reaches a saturation 
plateau. This saturation can be numerically controlled by requiring 
that $\langle 0|\varphi_j|0\rangle_i\sim R^{2\D_j}$, {\it i.e.} 
\EQ
{1\over 2\D_j}{d\ln\langle 0|\varphi_j|0\rangle_i\over d\ln~R}=1.
\EN
By using this procedure, several VEV for different deformations of 
the TIM were determined in \cite{GM1}. We have reproduced and 
confirmed the results of \cite{GM1} for the two most relevant 
perturbations $\sigma$ and $\epsilon$ of the model (see Table 
\ref{tvev}). In Table \ref{tvev}, we have also included 
the exact VEV extracted from ref.\,\cite{russian2} of the TIM 
perturbed by the thermal operator (for both high and low temperature 
phases) in order to test the feasibility of the numerical approach. 
As evident from this table, lower the dimension of the operator, 
better the accuracy of the method. This is easy to understand since 
the computation of the VEV on the cylinder is equivalent to compute 
the VEV at a finite--temperature, a situation analyzed in \cite{LeCMus}. 
The numerical determinations of the VEV's relative to the $\varphi_3$ 
and $\varphi_4$ deformations turn out to be quite inaccurate for 
the renormalization reasons discussed above ($\Delta_3 \sim 
\frac{1}{2}$ and $\Delta_4 > \frac{1}{2}$) but their exact 
values can be nevertheless extracted from the results of 
ref.\,\cite{russian2}.  

\subsection{Numerical Determination of the One--Particle 
Form Factors}

As discussed in Section 4, once the Form Factors of 
an operator are known, its correlation functions can be 
written as an infinite series over multi-particle states, 
eqs.\,(\ref{g_n}), (\ref{spectral}), and the restriction 
of these series to the first one--particle states 
already provides a reasonable approximation of their long 
distance behavior. These one--particle matrix elements
$\langle 0\mid\varphi_i(0)\mid A_k\rangle$ can be numerically 
determined along the lines followed for estimating the VEV of 
the various operators. Namely, one has to replace the infinite 
volume vacuum state $\langle 0 \mid$ with its components relative 
to the eigenvector of the truncated Hamiltonian, analogously for 
the vector relative to the one--particle state, and then use the 
matrix elements of the field $\varphi_i$ in the truncated basis. 
The resulting quantity finally needs to be multiplied by $\sqrt{m_k R}$ 
because the normalization of the one--particle states on a finite 
volume differs precisely for this factor from the one in the 
infinite volume. The quantity so determined, plotted versus the 
radius $R$ of the cylinder, presents in many cases a plateau in 
the region of the physical window which therefore provides its 
numerical estimation. As an example of such determination see 
Figure 3 where the matrix element 
$\langle 0\mid\varphi_1(0)\mid A_1\rangle_1$ is plotted 
as a function of $R$ in the case of the leading magnetic 
deformation of the TIM. A plateau is clearly reached for 
$R\ap 12$. We have performed this numerical calculation for 
all excited states under the two--particle threshold relative 
to the lowest mass gap for the first three deformations. The 
results are in Tables \ref{tff1}, \ref{tff2}, \ref{tff2m} and 
\ref{tff3}, in units of the opportune power of the coupling 
constants. As for the numerical determination of the VEV's, 
also in this case the lower the dimension of the operator, 
the accuracy of the method improves. Moreover, the numerical 
errors are usually larger for those matrix elements involving 
states which are closest to the threshold. 

Let us conclude this section with a general remark on the 
one--particle Form Factors.  We have already discussed that 
for the non--integrable deformations the knowledge of these 
matrix elements is crucial in obtaining at least a 
non--trivial estimate of the correlators and their determination 
necessarily passes through a numerical approach. However, even 
in the more favorable case of integrable theories, it may 
occur that the determination of the one--particle FF can be only 
obtained by a numerical approach. Despite the existence 
of a manageable set of recursive equations which link the various
$n$--particle Form Factors in the integrable models, the solutions of 
these recursive equations need an initial input which cannot be 
often obtained even by employing the cluster property of the Form Factors
\cite{DSC,cluster,AMVcluster}. Under this circumstance one has to 
necessarily resort to other methods for obtaining the one--particle 
FF's and the TCSA may help in this respect. An explicit example of the 
situation discussed above is provided for instance by the Form 
Factors of the operators $\varphi_1$ in the thermal deformation, 
as discussed in the Appendix \ref{ff}.

\subsection{Numerical Computation of the Susceptibilities}

The TCSA also allows a direct numerical estimation  
of the susceptibilities. Since they are defined as  
\EQ
\hat \Gamma_{jk}^i = 
\frac{\partial}{\partial g_k} \langle \varphi_j\rangle_i = 
\left. -\frac{\partial^2 f_i}{\partial g_k \partial g_j}\right|_{g_l=0} =
\Gamma_{jk}^i \,g_i^
{\frac{ \Delta_j +  \Delta_k - 1}
{1 - \Delta_i}} \,\,\, , 
\label{susc1}
\EN 
one needs to numerically evaluate the derivatives of the VEV's 
with respect to the different couplings. Hence, a small perturbation 
$g_k\int \limits_0^R dv \hat{\varphi}_k(w)$ is initially added
to the Hamiltonian (\ref{pert}), with the values of the coupling 
constant $g_k$ chosen in such a way to alter the spectrum of the 
unperturbed theory only of small percent. To express the final result 
in unit of $|g_i|^{{ \Delta_j + \Delta_k - 1\over 1 
-  \Delta_i}}$, it is convenient to write the coupling constant 
$g_k$ as 
\EQ
g_k=a_{ki}g_i^{{1-\Delta_j\over 1-\Delta_i}} \,\,\,.
\EN
The next step consists in computing the expectation value of 
$\langle \hat{\varphi}_j\rangle_{i+k}$ as in eq.\,(\ref{numvev}) 
by varying $a_{ki}$. As our typical sampling, we have considered 
$5-10$ different values of $a_{ki}$ and then we have extracted the
numerical estimates of various susceptibilities by a linear fit of 
the data in the physical window of the $R$ axis where the VEV 
presents a plateau. The data relative to the two strongest 
relevant deformations are in Tables \ref{tchi1}, \ref{tchi2} and 
\ref{tchi2m}. Their values are reasonably close to the ones 
obtained by the fluctuation--dissipation theorem or to their 
exact values, when available from the $\Delta$--theorem sum rule. 
The only exceptions are those relative to the susceptibilities 
relative to the $\varphi_4$ operator, where there is a $10\%$ 
mismatch. A non trivial and internal check of our estimates is 
given by the symmetrical relation $\Gamma_{jk}^i = \Gamma_{kj}^i$ 
shown by the data. As for similar calculations discussed above, 
this method seems however to be inappropriate for the higher 
dimension perturbations relative to the fields $\vp_3$ and $\vp_4$. 

\resection{The Four Relevant Perturbations of the TIM}

In this section, we will discuss in some details the field 
theories relative to each individual relevant deformation of 
the TIM, i.e. those associated to the actions  
\EQ
{\cal A}_i^{\pm} = {\cal A}_{CFT} \pm g_i \int d^2 x~ \vp_i(x) 
\,\,\,\,\, , \,\,\,\, i=1,\ldots,4
\label{action0}
\EN
The qualitative form of the effective potential relative to 
the different off--critical deformations of the TIM is shown 
in figure 4. In addition to the spectrum of the off--critical 
excitations, for each field theory we will present the main 
formulae involved in our estimation of the universal amplitude 
ratios of this model. Some aspects of the calculations of the 
correlation functions and the VEV are also discussed, refering 
to the appendices for all technical details. Here it is worth 
to comment on an interesting feature of the Conformal Perturbation 
Theory common to all deformations, i.e. the appearance of 
logarithmic terms both in some VEV and in the calculation of 
some susceptibilities. The origin of some of these term can be 
traced back to the existence of some peculiar resonance conditions 
involving the anomalous dimensions of this model. The first 
of them is given by 
\EQ
\Delta_1 = \Delta_3 + \Delta_4 -1 \,\,\,,
\label{resonance1}
\EN 
which is equivalent to say that the scale dimensionality of the 
coupling $g_1$ equals the product of $g_3$ and $g_4$, namely 
$g_1 \sim g_3 g_4$. It is easy to see that this resonance 
condition may influence the calculation of some susceptibilities. 
In fact, even though each individual off--critical field theory 
is defined by the one--coupling action (\ref{action0}), nevertheless 
the calculation of the susceptibilities $\hat\Gamma^i_{jk}$ requires 
to consider initially the multi--coupling action 
\EQ
{\cal A} = {\cal A}_i^{\pm} + 
g_j \int d^2 x~ \vp_j(x) 
+ g_k \int d^2 x~ \vp_k(x) \,\,\,,
\label{multi-action}
\EN
and to take the limit $g_j = g_k =0$ only at the end of the 
calculation. Hence, the resonance condition $g_1 \sim g_3 g_4$ 
may spoil the naive form of the above action (\ref{multi-action}) 
with the presence of additional terms. Explicit examples of 
this phenomenon are commented in the next sections. Another 
resonance condition which also influence some of the calculations 
is given by 
\EQ
\Delta_6 = 1 + \Delta_4 - \Delta_2 \,\,\,,
\label{resonance2}
\EN 
where $\Delta_6 = \frac{3}{2}$ is the anomalous dimension 
of the irrelevant field $\varepsilon"$. 

\subsection{The Magnetic $\varphi_1$ Deformation} 

This is the most relevant deformation of the TIM and the only
non--integrable one. Hence, most of the results relative to this 
deformation are obtained by the help of the numerical approach. 
The numerical analysis of the spectrum, first performed in 
\cite{LMC}, shows that there are two different one--particle 
states with mass ratio $m_2/m_1 \sim 1.61$ (see Figure 2). 
By setting the value of the first mass to be 1, its relationship 
with the coupling constant $g_1$ is numerically determined to be 
\EQ
m_1 = {\cal C}_1 \,
g_1^{{40\over 77}} \,
\approx 3.242...\,
g_1^{{40\over 77}} \,\,\,.
\EN
The VEV of the different fields have been numerically computed 
and their values are in Table \ref{tvevphi1}. By applying 
eq.\,(\ref{Gescond}), it is easy to check that there is no 
UV mixing of the operators for this deformation and therefore 
no need for their UV renormalization. In order to compute the 
various susceptibilities, we have decomposed 
$\int d^2 x ~\la \vp_i(x) \vp_j(0) \ra_1$ into 
the two integrals $I_1(R)$ and $I_2(R)$, as discussed in Section 4. 
The UV part of the correlator has been approximated by the short 
range expansion (\ref{correl}) with the employed values of 
$(C_{ij}^p)^{(1)}_1$ reported in Table \ref{tc1}. They were 
computed as explained in appendix B. For the IR part 
of the correlator, the non--integrable nature of this deformation 
forces us to truncate the spectral expansion to the one--particle 
contributions only
\EQ
\la \vp_i(x)\vp_j(0)\ra \,\approx\, 
\sum\limits_{l=1}^2 F_l^{\vp_i} F_l^{\vp_j} K_0(m_l|x|) \,\,\,.
\label{correlir}
\EN
The numerical estimation of the one--particle Form Factors $F_l$,
expressed in opportune units of $g_1$, can be found in Table \ref{tff1}. 

As a concrete example of the above procedure, let us consider 
the correlator $\la\vp_2(x)\vp_2(0)\ra_1$. Its UV expansion reads: 
\bea
\la \vp_2(r)\vp_2(0)\ra & = & r^{-4\D_2} \left[ 1+c_1 B_{41} 
\left({m_1r\over {\cal C}_1}\right)^2 + (C_{22}^1)^{(1)}_1  
B_{11} \left({m_1r\over {\cal C}_1}\right)^{2}\right.\nn + \\
&&+ \left. (C_{22}^3)^{(1)}_1 B_{31} 
\left({m_1r\over {\cal C}_1}\right)^{{14\over 5}} + 
{\cal O}((m_1r)^3) -(B_{21})^2
\left({mr\over {\cal C}_1}\right)^{4\D_2} \right]~.
\eea
The above expression is expected to provide an accurate approximation 
of the correlator up to $m_1r \sim 1$. Indeed, in a plot 
of the UV and the IR approximations of this correlator (Figure 5), 
a satisfactory overlap between the two curves is observed around 
$m_1 r \sim 1$: this makes us confident on the estimation of the 
susceptibility extracted by integrating the above correlator. 
The same situation occurs for the other susceptibilities and the 
final results are in good agreement with those extracted by the 
$\Delta$--theorem sum rule or their direct numerical estimation 
by TCSA. All these data are reported in Table \ref{tchi1}. 
The only exceptions consist in the calculation of the 
susceptibilities $\Gamma_{34}^1$ and $\Gamma_{44}^1$ for 
which some care is required due to the anomalous dimensions 
of the fields involved and to some subtleties in the 
conformal perturbation expansion of $C_{ij}^p(g;x)$. 
In these case, for instance, the naive numerical 
integration of the corresponding correlators cannot 
be performed because of their divergences at $r\to 0$.  
Let us first consider $\Gamma_{44}^1$. A natural way to 
regularise the integral $\int d^2 x ~\la \vp_4(x) \vp_4(0) 
\ra_1$ which near $r \to 0$ goes as $2\pi \int dr r^{-7/5}$, 
consists in introducing a UV cut-off $a$ so that its UV 
divergence is easily extracted  
\EQ
\Gamma_{44}^1 = {5\pi\over a^{{2\over 5}} } + 
\makebox{\rm finite \,part}\,\,\,.
\EN
Once this divergence is subtracted, the finite part of the integral, 
being cut-off independent, may be regarded as the {\it actual} 
regularized susceptibility. For the magnetic deformation its value is 
in Table \ref{tchi1} with $5-10\%$ of approximation and 
this quantity can be used later on in the evaluation of the universal 
ratios. The same strategy has been also adopted for the evaluation 
of the susceptibilities $\Gamma_{44}^i$ relative to the other 
deformations of the TIM since the divergence of the integral is 
simply due to the conformal properties of the field $\varphi_4(x)$ 
and does not depend on the particular deformation considered.  

The situation is different for $\Gamma_{34}^1$. In this 
case, by using the operator product expansion, it is easy to 
see that the integral $\int d^2 x ~\la \vp_3(x) \vp_4(0) \ra_1$ 
behaves in the UV region as 
\EQ 
\int_a^R \frac{dr}{r} \,\,\, ,
\EN
and therefore it presents a logarithmic divergence
\bea
\Gamma_{34}^1 & = & {3\pi\over 2} \,B_{11} \ln\left(
{\Lambda\over a}\right) + G\nn \\  
& = & -{3\pi\over 2(2-2\D_1)} \,B_{11}
\ln\left({g_1\over g_1^0}\right) + G \,\,\, , 
\label{gamma34}
\eea 
where $G$ is the finite part. When $B_{11}$ is different from $0$, 
$G$ is not uniquely defined since it varies by changing the value 
of $g_1^0$. Hence, contrary to the previous case, the finite part of 
the integral cannot be used to define universal ratios although  
the amplitude in front of the logarithmic term is an unambiguous 
quantity in QFT which may enter universal combinations. It is worth 
pointing out that this situation is not peculiar of the TIM 
but it is already familiar in the context of studying the specific 
heat dependence in the standard two--dimensional Ising model (see, for
instance \cite{DelfinoIsing}). Obviously the above considerations apply 
for all the susceptibilities $\Gamma_{34}^i$ relative to the 
other deformations of the TIM.

\subsection{The Thermal $\vp_2$ Deformation}
\label{vp2pert}
This is an integrable deformation of the TIM \cite{MC,FZ}. When 
$g_2 > 0$, the coupling to the thermal field moves the TIM in 
its high--temperature phase, where a unique $Z_2$ symmetric 
vacuum state is present. When $g_2 < 0$, we reach the 
low--temperature phase of the model, where there is a 
spontaneously breaking of the $Z_2$ spin symmetry and 
therefore two degenerate symmetric vacua. In the high--temperature 
phase there are ordinary massive particle excitations whereas 
in the low--temperature phase there are kink excitations and bound 
states thereof. The two phases are related each to the other by 
a duality transformation. The off--critical model possesses 
higher conserved charges whose spins are $s=1,5,7,9,11,13,17$ 
(module $18$), i.e. the Coxeter exponents of the exceptional 
algebra $E_7$. The integrable structure of this deformation 
originates from the conformal decomposition (\ref{E7WZWM}) together 
with the pairing of the energy operator $\epsilon(x)$ with the 
adjoint representation of the WZW model on $(E_7)_2$ \cite{EY}. 
The existence of an infinite number of conservation laws 
implies the elasticity and factorization of the scattering amplitudes. 
These amplitudes have been computed in \cite{MC,FZ} and their 
concise expressions can be found in Table 2 of ref.\,\cite{AMV}.  
The exact mass spectrum of the excitations can 
be extracted from the pole structure of the $S$ matrices (see 
Table \ref{tspectrum}). With respect to the $Z_2$ spin symmetry 
of the model, in the high--temperature phase there are three 
$Z_2$ odd particle states (the ones relative to the masses 
$m_1$, $m_3$ and $m_6$) and four $Z_2$ even (those relative 
to the masses $m_2$, $m_4$, $m_5$ and $m_7$). In the 
low--temperature phase, the three $Z_2$ odd particle 
states become kink excitations interpolating between 
the two degenerate ground states whereas the four $Z_2$ 
even ones play the role of breather states. This leads, in 
particular, to an interesting prediction on the universal 
ratio of the correlation lengths {\em above} and {\em below}
 the critical temperature. In fact, if the correlation length 
is defined according to the leading exponential falling off 
of the spin--spin connected correlation function in the 
limit $\mid x\mid \gg \xi^{\pm}$  
\EQ
\langle 0 \mid \vp_1(x) \vp_1(0) \mid 0\rangle_c^{\pm} 
\sim \exp\left(-\frac{\mid x\mid}{\xi^{\pm}}\right) 
\,\,\, ,
\label{falling}
\EN 
(where the indices $\pm$ refer to the high and low temperature 
phases respectively), from the $Z_2$ symmetry property of the 
$\sigma$ field, the self--duality of the model and the 
spectral representation of the above correlator we have 
\EQ
\frac{\xi^+}{\xi^-} = \frac{m_2}{m_1} = 2 \cos\frac{5\pi}{18} =
1.28557...
\label{xiuniv}
\EN  

For this deformation the relationship between the mass--gap $m_1$ 
and the coupling constant $g_2$ is given by $m_1 = {\cal C}_2\, 
g_2^{{5\over 9}}$ where \cite{fateev}  
\bea {\cal C}_2   &=&
\left({2{\Gamma({2\over9})}\over{\Gamma({2\over3})}
{\Gamma({5\over9})}}\right)
\left({4\pi^2\Gamma({2\over5}){\Gamma({4\over5})}^3 
\over{\Gamma({1\over 5})}^3{\Gamma({3\over 5})} } 
\right)^{5/18} =  3.7453728362 \dots
\eea               
The VEV's of all relevant operators, both in the high or in the low 
temperature phases, have been computed in \cite{russian2} and their 
values successfully compared with their numerical determination 
\cite{GM1}: the expressions of the expectation values of the 
operators $\sigma$ and $\epsilon$ are finite, whereas those of 
the operators $\sigma'$ and $\epsilon'$ are {\it naively} divergent 
(see eq.\,(\ref{VEVepsilon})) and need therefore a regularization 
(see Appendix C). They are presented in Table \ref{tvev}, in opportune 
units of the associate power of the coupling constant. Concerning the 
UV properties of the fields, eq.\,(\ref{Gescond}) predicts that there 
is no UV mixing of the operators for this deformation and therefore 
no need to implement their UV renormalization. 

Let us now turn the attention to the computation of the various 
susceptibilities by following the strategy explained in Section 
4. The values of the coefficients $(C_{ij}^{p})^{(1)}_2$ entering 
the UV expansion of the correlators are in Table \ref{tc2}. As 
a concrete example of these calculations, let us discuss the 
susceptibility amplitude $\Gamma_{11}^2$ relative to the correlator 
$\la \vp_1(x)\vp_1(0)\ra_2$. Its UV expansion reads: 
\bea
\la \vp_1(r)\vp_1(0)\ra_2 & = & r^{-4\D_1} \left[ 1 + c_5 B_{22}
\left({mr\over {\cal C}_2}\right)^{{1\over 5}}
+c_7 B_{42} \left({mr\over {\cal C}_2}\right)^{{6\over 5}}+
(C_{11}^{0})^{(1)}_2  
\left({mr\over {\cal C}_2}\right)^{{9\over 5}}\right.\nn\\
&&+ \left. (C_{11}^{2})^{(1)}_1 B_{22} \left({mr\over {\cal C}_2}
\right)^{2}
-(B_{12})^2 \left({mr\over {\cal C}_2}\right)^{4\D_1} +
{\cal O}((mr)^3)\right]\,\,\,.
\eea
 
\no The IR part of the correlator is approximated by taking initially  
into account the first four one--particle states of the spectral 
expansion of the correlator (see eq. (\ref{spectral})) 
\EQ
\la \vp_i(x)\vp_j(0)\ra \approx \sum\limits_{l=1}^4 
F_l^{\vp_i} F_l^{\vp_j} K_0(m_l|x|) \,\,\,.
\EN
The numerical values of the needed one--particle Form Factors are 
in Tables \ref{tff2} and \ref{tff2m}, relative to the high and low 
temperature phases respectively. Although these matrix elements are 
already able to reproduce with a reasonable accuracy the infrared part 
of the correlator, due to integrability of this deformation some 
two--particle Form Factors are also available and therefore they can 
be used to improve the estimation of the correlators. Their calculations, 
together with some subtleties which occur in this case, are discussed 
in Appendix D. The Form Factors of the operator $\epsilon(x)$, which 
plays the role of the trace of the stress--energy tensor for this 
deformation, were computed in ref.\,\cite{AMV}. 

The above strategy has been applied for the estimation of all 
the correlators. An overlap between the UV and IR approximations 
of the correlators has been usually observed in the region $m_1 r 
\sim 1$, which may be regarded as a consistent check of our 
approach. Such overlap is shown in Figure 6 for 
the correlator $\la\vp_1(x)\vp_1(0)\ra_2$.

In closing this subsection, some comments are in order for 
Table \ref{tc2} which collects the values of the first corrections 
to the structure constants in the thermal deformation. Three of 
these coefficients contain some logarithmic dependence which 
however do not particularly influence the numerical approximation of 
the short--distance of the correlators since these are
higher--order corrections. The presence of these logarithms 
are due to a resonance phenomenon between the conformal families, 
i.e. it occurs when the conformal dimensions of two operators differ 
for an integer number of times $1-\Delta$, where $\Delta$ is 
the dimension of the perturbing operator. This situation is 
encountered here because $$\D_{6}-\D_{4} = (1-\D_{2}).$$ Hence, 
we have for instance $(C_{11}^4)^{(1)}_2(mr) = a_1(\ln(mr)+b_1(m))$ 
with the coefficient $a_1$ that can be computed by using the methods
explained in Appendix 2. Since its algebraic expression is rather 
cumbersome, we prefer to report here just its numerical value 
$a_1 = -0.1510653\dots$. While $a_1$ is an unambigous quantity in 
QFT, the other term $b_1(m)$, on the contrary, is a scale dependent 
quantity. Similar resonance phenomena are also encountered in 
the sub--leading magnetic deformation of the TIM, which is the 
subject of the next subsection, where they play a more important 
role since they contribute to the lower orders in the UV expansion.

\subsection{The Sub--leading Magnetic $\vp_3$ Deformation}
\label{vp3pert}

This is an integrable perturbation \cite{Smirnov12,CKM}. 
It is generated by the less relevant magnetic field and 
obviously breaks the $Z_2$ spin symmetry of the critical 
point since the field $\vp_3$ is a $Z_2$ odd operator. 
The resulting massive theory presents some interesting 
features, as first outlined in ref.\,\cite{LMC}. The theory 
presents two degenerate albeit asymmetrical vacua (denoted 
by $|0_2\ra$ and  $|0_4\ra$). There are two massive kink 
excitations of mass $m$ and one breather bound state thereof with 
the same mass. The exact $S$--matrix of this theory has been 
computed and analyzed in \cite{CKM}. The relationship between the 
coupling constant $g_3$ and the mass gap of the theory is provided 
by $m = {\cal C}_3 g_3^{{8\over 9}}$ with the constant ${\cal C}_3$ 
given by \cite{fateev} 
\bea
{\cal C}_3 & = &{ \sqrt3\,\Gamma\big({1\over 3}\big)\, 
\Gamma\big({5\over 9}\big) \over \pi\, \Gamma\big(
{8\over 9}\big)}\ \biggl[\,  {\pi^2\,  \Gamma^2\big({11 \over 16}
\big)\, \Gamma\big({1\over 4}\big) \over \Gamma^2\big({5\over 16}
\big)\, \Gamma\big({3\over 4}\big) }\, \biggr]^{{4\over 9}} = 
4.92779064\dots    
\eea
The VEV's of relevant operators $\vp_1$, $\vp_2$, $\vp_3$ have been
exactly computed  in \cite{russian2}. The expression of the VEV's 
$\la 0_s| \vp_4|0_s\ra$, $s=2,4$ given by eq.\,(\ref{VEVsigma'}) 
are {\it a priori} divergent and need to be regularized. As shown 
in Appendix C and also discussed below, they present a logarithmic 
dependence on the coupling constant $g_3$. The complete set of VEV,
expressed in unit of the appropriate power of the coupling constant, 
are in Table \ref{tvev3}. Notice that, according to the formula 
(\ref{Gescond}), the only field which requires an ultraviolet 
renormalization is precisely the field $\vp_4(x)$ which mixes 
(logarithmically) with the field $\vp_1(x)$ under perturbation 
theory. 

The application of the $\Delta$--theorem sum rule allows the 
exact determination of the four susceptibilities $\Gamma_{3j}^3$  
(as already discussed in Section 6.1, the susceptibility $\Gamma_{34}^3$ 
contains a logarithmic dependence on the coupling constant $g_3$). 
Also for this deformation we have followed the general strategy 
explained in Section 4, i.e. we have first pursued the matching 
between the UV and the IR expansions of the correlators and then 
we have performed their integration for extracting the susceptibilities. 
Let us discuss the correlator $\langle\vp_1(r)\vp_1(0)\rangle_3$ 
to exemplify some new difficulties arising in the perturbative 
evaluation. Its UV expansion is given by 
\bea
\langle\vp_1(r)\vp_1(0)\rangle_3 &=&r^{-4\Delta_{1}} \left(
1+c_5 B_{23} \left({mr\over {\cal C}_3}\right)^{{1\over 5}}
+c_7 B_{43} \left({mr\over {\cal C}_3}\right)^{{6\over 5}}
+(C_{11}^1)^{(1)}_3 B_{13} \left({mr\over {\cal C}_3}\right)^{{6\over 5}}
\right. \nn\\  
&+&  \left. (C_{11}^{3})^{(1)}_3(r) B_{33} \left({mr\over
{\cal C}_3}\right)^{2} + \dots
-(B_{13})^2 \left({mr\over {\cal C}_3}\right)^{4\Delta_{1}} 
\right) \,\,\,.
\eea
There are two different sources of problems in this expansion. 
First of all, the VEV $ \la \vp_4 \ra_3$ contains a logarithm 
dependence on the coupling and a UV regularization is needed. 
By using eq.\,(\ref{gamma34}) and the $\Delta$--theorem 
sum rule  
\EQ
B_{43}=-{(1-\Delta_{3})\over \D_{4}}\,\Ga_{34}^3~,
\EN
it can be easily shown that 
\EQ
B_{43} = -{3\pi(1-\D_3)\over 2 \D_4}  B_{13} 
\left(\ln{\Lambda\over \epsilon} + G_3 \left({\Lambda\over \epsilon}\right)~\right) 
= {3\pi\over 4 \D_4} B_{13} \left(\ln{g_3\over g_3^0} + 
\tilde{G}_3 \left(
{g_3\over g_3^0}\right)~\right) \,\,\, , 
\EN
where $\epsilon$ is some arbitrary UV cut-off and $\Lambda\sim \xi_3^0$. 
Secondly, the coefficient $(C_{11}^1)^{(1)}_3$ contains some logarithmic 
divergences due to resonance problems already encountered in the 
thermal perturbation and cannot be calculated by using the Mellin 
transformation method in the usual manner. However, the Mellin 
transformation method can be properly generalized to obtain the 
correct logarithmic term, as we show in the following.
$(C_{11}^1)^{(1)}_3(r)$ is defined as usual by the regularized integral 
(notice that we have restored the dependence on $r$)
\EQ 
(C_{11}^1)^{(1)}_3(r)=-\int' d^2 z \la
\vp_1(\infty)\vp_1(r)  \vp_3(z)\vp_1(0)\ra \,\,\,.
\EN
Using notation of Appendix A, $(C_{11}^1)^{(1)}_3(r)$ may be regarded 
as the {\it regularized} limit $s\rightarrow 0$ of   
\bea 
\tilde{G}(2-s;r)&=&\int d^2z \,|z|^{-s}\,\la \vp_1(\infty)\vp_1(r) 
\vp_3(z)\vp_1(0)\ra \nonumber \\ 
&=&r^{-s}\int \int d^2z d^2u |u|^{2a}|u-1|^{2b}
|u-z|^{2c}|z|^{2d-s}|z-1|^{2e}
\label{tildeGr}
\eea
where the coefficients $a,b,c,d,e$ are expressed in terms of the 
weights by the Coulomb gas formalism and $s$ is the Mellin transformation 
parameter. After putting $r=1$ in the previous equation, we have 
\EQ 
\tilde{G}(2-s;1)=-\frac{{3\pi\over 8} c_1}{s} +{\cal O}(1) \,\,\,,
\EN
i.e. there is a first order pole which does not allow us to calculate 
$(C_{11}^1)^{(1)}_3$ in the usual manner. However, by multiplying the 
previous expression with the expansion (see
(\ref{expcorrectionexp})) 
\EQ
r^{-s} = m^s(mr)^{-s} = [1-s\ln mr + {\cal O}(s^2)] \,\,\,,
\EN
where m is an arbitrary mass scale, we obtain the expansion in $s$
\EQ
\tilde{G}(2-s;r)=-\frac{{3\pi\over 8} c_1}{s} + {3\pi\over 8} c_1 
\ln mr + {\cal O}(1) \,\,\,.
\EN
Hence, $(C_{11}^1)^{(1)}_3(r)$ may be taken as the regularized version 
of the previous expression, i.e. the one obtained by discarding the 
simple pole divergence 
\EQ
(C_{11}^1)^{(1)}_3(r) = {3\pi\over 8} c_1 (\ln (mr)+ \makebox{\rm const})
\,\,\, ,
\EN
where the constant term clearly depends on the scale $m$. By looking 
at eq.\,(\ref{tildeGr}) and by taking into account the dependence of
$a,b,c,d,e$ on the weights of the primary fields, it is easy to see 
that a shift of the parameter $\xi$, defined in eq.\,(\ref{xieta}), 
$\xi=\frac{p}{p'-p}\rightarrow \xi+\epsilon$ would have
led to the same result, with $\epsilon$ playing the r\^ole of $s$. 
The two terms in $\left({mr\over {\cal C}_3}\right)^{{6\over 5}}$ lead 
in the correlator to 
\EQ
{3\pi\over 8} c_1 B_{13}\left({mr\over {\cal C}_3}\right)^{{6\over 5}}
(\ln(mr) + A(m)) \,\,\,,
\EN
where $A$ is a cut-off dependent quantity which can be determined 
by requiring the matching between the infrared expansion of the 
correlator with its UV part. The infrared part is approximated by the 
one--particle Form Factors of the magnetic and thermal operators 
(which are reported in Table \ref{tff3}). There is a non-trivial 
consistency check of the above procedure. Indeed, the correlation 
function can be computed in the two different vacua $|0_2\ra$ 
and $|0_4\ra$. Once the quantity $A$ has been fixed by computing 
the correlator $\la 0_2|\vp_1(r)\vp_1(0)|0_2\ra$, the same $A$ 
should also work for the other correlator $\la 0_4|\vp_1(r)
\vp_1(0)|0_4\ra$, as it is indeed the case. 

The above procedure has been also employed for the correlators 
$\la\vp_2(r)\vp_2(0)\ra_3$ and for $\la\vp_1(r)\vp_2(0)\ra_3$. For 
this deformation, however, we were unable to reach any definite 
result on those correlators involving the $\vp_4$ operator because 
its one--particle FF were completely inaccurate, thus preventing a 
reliable evaluation of the infrared part of the correlators. The results 
for all the susceptibilities we were able to compute for this deformation 
are in Table \ref{tchi3}.

\subsection{The Vacancy Density $\vp_4$ Deformation}
\label{vp4pert}

The field theories originated by the perturbation $t(x) =\vp_4(x)$ 
with $g_4 > 0$ and $g_4 < 0$ are both integrable \cite{Zamthree}. 
The most elegant way to get an insight on these quantum field 
theories is to use the supersymmetric formulation of the TIM 
\cite{ZamLG,Kastor,Zamthree}. Since the field $t(x)$ is the top 
component of the superfield (\ref{superfield}), the off--critical 
dynamics may be described by the action 
\EQ
{\cal A} = \int d^2x \, d^2\theta \,
\left[ \frac{1}{2} {\cal D} {\cal N} \,
{\bar {\cal D}} {\cal N} 
+ {\cal N}^3 + g_4 \,{\cal N} \right]\,\,\, . 
\label{offsuper}
\EN 
After eliminating $t(x)$ by its algebraic equation of motion, 
the interaction terms of the above lagrangian are given by 
\EQ
\overline\psi \psi \,\epsilon + \frac{1}{2} \left(
\frac{1}{2} \epsilon^2 + g_4\right)^2 \,\,\,. 
\label{interaction}
\EN 
Hence, for $g_4 > 0$ the ground state energy is nonzero and 
supersymmetry is spontaneously broken: the scalar field acquires 
a mass whereas the fermionic field remains massless and plays 
the role of goldstino. This is nothing but the Majorana 
fermionic field of the familiar two--dimensional Ising model 
which is in fact the ending point of the massless Renormalization 
Group flow originating from the $g_4 > 0$ deformation \cite{Kastor}. 
On the massless Majorana fermion of the Ising model supersymmetry is 
implemented non--linearly. The exact massless $S$--matrix has 
been computed in \cite{Zammassless} and the related massless 
Form Factors in \cite{DMSmassless}. The crossover phenomena, 
where the exponents characterizing the leading singularity 
of an observable change from its tricritical to its critical 
value, have been studied in \cite{DMSmassless}. Along this 
flow, the conformal dimension of the magnetization field 
changes from its tricritical value $\frac{3}{80}$ to 
$\frac{1}{16}$ of the Ising fixed point, the conformal 
dimension of the energy operator varies from $\frac{1}{10}$ 
to $\frac{1}{2}$ and finally the conformal dimension of the 
vacancy density operator changes from $\frac{6}{10}$ to $2$, 
since this operator becomes -- at the end of the Renormalization 
Group flow -- a descendent of the Identity operator. The 
sub--leading magnetization operator also becomes at the end 
of the RG flow a descendent of the magnetization field of the 
Ising model. For the massless nature of this theory we will 
not compute the relative universal amplitude ratios.
 
For $g_4 < 0$ the ground state energy vanishes therefore 
supersymmetry is unbroken and both the scalar and the 
fermion fields become massive. As argued in \cite{Zamthree}, 
to describe the dynamics of the system it is more convenient 
to adopt the usual Landau--Ginzburg potential in terms of the 
order parameter $\sigma(x)$. This potential presents three--fold 
degenerate vacua, labelled by $|0_{-1}\ra, ~|0_0\ra, ~|0_{+1}\ra$, 
where $|0_0\ra$ corresponds to the disordered vacuum and 
$|0_{\pm 1}\ra$ to the two disordered vacua, symmetrically placed 
with respect to the origin (see Figure 4). The elementary 
excitations in this phase are massive kinks, which interpolate 
between the ordered vacuum and the disordered vacua and are 
denoted by $K_{0\pm 1}$ or $K_{\pm 1 0}$. The associated scattering 
theory has been discussed in \cite{Zamthree,delfino99}.

Let us consider the amplitude ratios for this deformation. The VEV 
of $\la \vp_1\ra_4$ and $\la \vp_2\ra_4$ can be directly computed by 
using eq.\,(\ref{VEVepsilon'}), since the integral converges. The
VEV's of $\vp_3$ contains a logarithm divergence and is calculated 
in Appendix C. This is the mirror situation encountered in Section 
\ref{vp3pert} for the VEV's $\la \vp_4\ra_3$ and therefore it can 
be fixed as in the previous subsection: 
\EQ
\la \vp_3 \ra_4  = {3\pi\over 4 \D_3} B_{14} \left(
\ln{g_4\over g_4^0} + \makebox{\rm const} \right)~. 
\EN 
Finally, by  supersymmetry we expect $\la \vp_4\ra_4 = 0$, because 
the field $\vp_4(x)$  plays the role of the trace of the stress--energy 
tensor for this  deformation and indeed this is in agreement with
formula (\ref{VEVepsilon'}). The values of the VEV may be found 
in Table \ref{tvev4}.

Concerning the UV properties of the theory, there are two
fields which need renormalization. The first is the sub--leading 
magnetization $\sigma'(x)$, which mixes logarithmically 
with the magnetization field $\sigma(x)$. The second is 
the vacancy density field $t(x)$, which mixes with the 
energy density $\epsilon(x)$. To avoid the difficulties 
related to the mixing induced by renormalization, for 
this deformation it is convenient to rely only on the 
Form Factor expansion for estimating the correlation functions. 
In fact, all fields, except the sub--leading magnetization 
$\sigma'(x)$, turn out to be uniquely identified by 
their symmetry properties. Moreover, the spectral series 
based on the Form Factor are also able in this case 
to capture successfully the ultraviolet behavior of the 
correlators \cite{delfino99}. Hence, for this deformation 
the integrals (\ref{I1I2}) will be estimated only in terms of 
$I_2(0)$, i.e. $I \simeq I_2(0)$. The two--particle matrix 
elements on the kink states for the operators $\vp_1, ~\vp_2$ 
and $\vp_4$ have been calculated in \cite{delfino99} by using 
a mapping of the TIM onto the dilute q-state Potts model, 
with $q=2$. All these operators are coupled to states with 
zero topological charge, i.e. to those multi--kink states 
which begin and end to the the same vacuum. There are two 
types of such two--kink Form Factors, depending whether 
the vacuum\footnote{Note that we use a different notation 
for the vacua compared to the one used in \cite{russian2}. 
These notations enable us to specify more clearly the disorder 
vacuum $|0_0\ra$ (with zero magnetization) and the two ordered 
vacua $|0_{\pm 1}\ra$ (with $\pm 1$ magnetization).} 
is $|0_0\ra$ or $|0_{\pm 1}
\ra$ 
\bea
F_{0\pm 1}^{\vp_i}(\theta_1-\theta_2) &\equiv&\la 0_0|\vp_i|
K_{0\pm 1}(\theta_1)K_{\pm 1 0}(\theta_2)|0_0\ra \,\,\,;\\
F_{\pm 1 0}^{\vp_i}(\theta_1-\theta_2) &\equiv&\la 0_{\pm 1}|
\vp_i|K_{\pm 10}(\theta_1)K_{0\pm 1}(\theta_2)|0_{\pm 1}\ra 
\,\,\,,
\eea
where $\vp_i$ designs the operator under consideration.
The Form Factors can be conveniently parameterized as  
\bea
&&F^{\vp_i}_{0\pm 1}(\th)=F^{\vp_i}_-(\th)\,\,,\hspace{1cm} i=2,4 \\
&&F^{\vp_i}_{\pm 10}(\th)=F^{\vp_i}_+(\th)\,\,,\hspace{1cm} i=2,4\\
&&F^{\vp_1}_{0\pm 1}(\th)=\pm F^{\vp_1}_-(\th)\,\,,\\
&&F^{\vp_1}_{\pm 10}(\th)=\pm F^{\vp_1}_+(\th)\,\,\,,
\eea
and their expressions are given by \cite{delfino99} 
\bea
&& F^{\vp_4}_{\pm}(\th)=-i\pi m^2e^{\pm\frac{\gamma}{2}(\pi+i\th)}\,
\frac{\cosh\frac{\th}{2}}{\sinh\frac{1}{4}(\th-i\pi)}\,F_0(\th)\,\,,
\\
&& F^{\vp_2}_{\pm}(\th)=\pm i(\la 0_{\pm 1}|\vp_2|0_{\pm 1}
\ra-\la 0_{0}|\vp_2|0_{0}\ra    )\,
\frac{e^{\pm\frac{\gamma}{2}(\pi+i\th)}}
{4\sinh\frac{1}{4}(\th-i\pi)}\, F_0(\th)\,\,,\\
&& F^{\vp_1}_{\pm}(\th)=\mp\frac{\la 0_{+ 1}|\vp_1|0_{+1}\ra}
{2\Upsilon_+(i\pi)}\,
\frac{e^{\pm\frac{\gamma}{2}(\pi+i\th)}}{\cosh\frac{\th}{2}}\,
\Upsilon_\pm(\th)F_0(\th)\,\,,
\eea
with
\[
F_0(\th)=-i\sinh\frac{\th}{2}\,\exp\left\{\int_0^\infty-\frac{dx}{x}\,
\frac{ \sinh(\frac{3x}{2})}{\sinh 2x\cosh\frac{x}{2}}\,
\frac{\sin^2(i\pi-\th)\frac{x}{2\pi}}{\sinh x}\right\}\,\,,
\]
\[
\Upsilon_+(\th)=\exp\left\{2\int_0^\infty\frac{dx}{x}\,
\frac{\sinh(\frac{p}{2}-1)x}{\sinh\frac{px}{2}}\,
\frac{\sin^2(2i\pi-\th)\frac{x}{2\pi}}{\sinh 2x}\right\}\,\,,
\]
\[
\Upsilon_-(\th)=\Upsilon_+(\th+2i\pi)\,\,,
\]
\[
\gamma={1\over 2\pi}\ln 2\,\,\,.
\]
With the knowledge of the first Form Factors, the spectral 
representations of the correlation functions involving 
the fields $\vp_1(x)$, $\vp_2(x)$ and $\vp_4(x)$ are 
approximated by 
\bea
\langle 0_0|\vp_i(x)\vp_j(0)|0_0\rangle \simeq \sum_{k=\pm}
\int_{\th_1>\th_2}\frac{d\th_1}{2\pi}\frac{d\th_2}{2\pi}
F^{\vp_i}_{0\pm 1}(\th_1-\th_2)F^{\vp_j}_{0\pm 1}(\th_2-\th_1)
\,e^{-|x|E_2} 
\,\,\, ; \label{approx1}\\
\langle 0_{\pm 1}|\vp_i(x)\vp_j(0)|0_{\pm 1}\rangle \simeq 
\int_{\th_1>\th_2}\frac{d\th_1}{2\pi}\frac{d\th_2}{2\pi}
F^{\vp_i}_{\pm 10}(\th_1-\th_2)F^{\vp_j}_{\pm 10}(\th_2-\th_1)\,
e^{-|x|E_2} 
\,\,\,,\label{approx2}
\eea
where $i,j=1,2,4$ and $E_2=m(\cosh\th_1+\cosh\th_2)$ is the 
energy of the two-kink asymptotic state. By integrating the 
above expressions, one can obtain the associated amplitudes. 
The convergence of the above spectral series has been successfully 
checked against the $\Delta$--theorem sum rule (when this applies)  
\cite{delfino99}. Our results for the amplitudes are in Table 
\ref{tchi4}. 

\resection{Conclusions}

In this paper we have exploited some powerful techniques of 
Quantum Field Theory in order to compute an ample set of universal 
amplitude ratios for the scaling region of the two--dimensional 
Tricritical Ising Model. The determination of the thermodynamical 
amplitudes entering the universal ratios has been obtained 
by combining exact non-perturbative results coming from CFT (UV 
theory) and from Scattering Theory (IR theory). More specifically, 
we have used the ultraviolet data provided by CFT for setting up 
an Operator Product Expansion (and computing the first order 
approximation to the structure constants), and for finding 
eigenvalues and eigenvectors of the off-critical hamiltonian 
by a numerical approach. In addition, we have used non--perturbative 
approaches related to the integrability of several deformations 
of the model to obtain important infrared data, as for instance 
the exact values of the Vacuum Expectation Values of the order 
parameters, the exact spectra of the massive excitations and the 
first Form Factors. A judicious use of the ultraviolet and 
infrared properties of the various fields which span the scaling 
region of the model have allowed us to reduce significantly 
the analytic efforts necessary to compute the universal ratios. 
Some of these quantities can be found in Tables 22--26\footnote{
In the calculation of the universal ratios, we have used 
the exact values of the susceptibilities when available 
from the $\Delta$--theorem sum rule, otherwise we have 
used the arithmetic mean of their determinations obtained 
by the fluctuation--dissipation theorem and the numerical 
TCSA.}. As already pointed out in the text, this large set 
of quantities may be quite useful for an experimental 
investigation of the critical properties of this class of 
universality and we hope that the results presented in this 
paper may stimulate such experimental activity. It would be equally 
interesting to extend the theoretical approach discussed here to 
other two--dimensional models in order to reach a full control 
of the classes of universalities of low--dimensional systems.

\vspace{5mm}{\em Acknowledgements}. We are pleased to thank 
Michele Caselle, Alexander and Alexei Zamolodchikov for useful 
discussions. This work has been done under partial support of the 
EC TMR Programme ERBFMRXCT960012 {\em Integrability, 
non-per\-turba\-tive effects and symmetry in 
Quantum Field Theories}. D.F. would like to thank I.N.F.N.
for a fellowship and S.I.S.S.A. for hospitality.

\newpage

\appendix

\section{Mellin regularization scheme}
The aim of this appendix is to discuss a regularization of 
the integrals (\ref{firstorderc}). They can be written as  
\EQ
C_{ij}^{p(1)} = -\int ' d^2w \, g(w,\bar{w})    \,\,\,,
\label{analytic}
\EN
with 
\EQ
g(w,\bar{w}) \equiv  \langle \tilde{A}^p(\infty) 
\tilde{\Phi}(w) \tilde{\varphi}_i(1) \tilde{\varphi}_j(0)
\rangle_{CFT}.
\EN
A regularized version of the divergent integrals (\ref{firstorderc}) 
is provided by the following function of the large distance cut-off 
$R \equiv 1/a$
\EQ
I(a) = -\int d^2w \,\Theta(a|w|) \,g(w,\bar{w}) \,\,\,, 
\label{regularization}
\EN
where the cut--off function $\Theta(t)$ has a fast 
decreasing behavior at $t\rightarrow +\infty$ to make 
the integral convergent and is equal to $1$ for $t\rightarrow 
0^+$. Of course $I(a)$ diverges for $a\rightarrow 0^+$ in 
a way that depends on the particular choice of $\Theta(t)$, 
but it converges to a finite value for $a\rightarrow +\infty$ 
thanks to the fast decreasing behavior of the cut--off function. 
The finite part of the integral (\ref{regularization}) is independent 
of the parameter $a$ and furnishes its natural regularization. 
It coincides with the analytic continuation of the integral 
(\ref{analytic}) from those regions of the conformal weights for
which it converges. The finite part  of the integral can be explicitly 
calculated by means of the Mellin transform of the complex function 
$I(a)$ defined by  
\EQ
\tilde{I}(s) = \int_0^\infty\frac{da}{a} a^s I(a) \,\,\,.
\EN 
In fact, when the Mellin transform have simple poles, it 
provides the asymptotic expansion in powers of $a$ of the 
function $I(a)$ according to the formula 
\EQ  
I(a) = \sum_i {\makebox {\rm Res}} [a^{-s}\tilde{I}(s)]_{s=s_i} \,\,\,,  
\EN
where the sum runs over the poles. In our cases the above 
sum is finite since the theory presents only a finite number 
of IR divergent terms. The finite part of $I(a)$ is therefore 
given by 
\EQ
I_0 = {\makebox {\rm Res}} [\tilde{I}(s)]_{s=0} = 
\lim_{s\rightarrow 0}s\tilde{I}(s) \,\,\,.
\EN  
Finally, we also need the following theorem on convolution: 
if the  function $I(a)$ has the form of a convolution (as in 
equation (\ref{regularization})), then its Mellin transform 
is given by
\EQ
\tilde{I}(s) = -\tilde{\Theta}(s) \, \tilde{G}(2-s) \,\,\, ,
\EN
where $\tilde{\Theta}(s)$ is the Mellin transform of $\Theta(t)$ and
$\tilde{G}(s)$ may be considered as the Mellin transform 
of the angular integral of $g(|w|,arg(w))$ with respect to the radial
coordinate $|w|$ 
\EQ
\tilde{G}(2-s) = \int d^2w \,|w|^{-s}\,g(w,\bar{w})\,\,\,.
\label{tildeG}
\EN
In our calculation we have used for $\Theta(t)$ the function  
\EQ
\Theta(t) = e^{-t} \,\,\, ,
\EN
whose Mellin transformation is the Gamma function
\EQ
\tilde{\Theta}(s) = \Gamma(s)\,\,\,.
\EN
For the calculation of $\tilde{G}(2-s)$ we refer the reader 
to the next appendix. 

\section{ Calculation of the $(C_{ij}^{p(1)})_k$}  

In this appendix, we show how to compute the first correction to 
the structure constants, i.e. the finite part of  
\EQ
(C_{ij}^p)^{(1)}_k= \int\limits^{'} \la
\vp_p(\infty) \Phi_k(w)\vp_j(1)\vp_i(0)\ra \,d^2w\,\,\,.  
\EN
The conformal four--point correlation functions
entering the integral may be computed by means of the 
modified Coulomb Gas methods \cite{DF}. In this approach 
the central charge $c$ is parameterized by 
\bea
c = 1-24\alpha_0^2~~
&;&~~\alpha_{\pm}=\alpha_0\pm\sqrt{\alpha_0^2+1};\\
\alpha_+\alpha_- &=&-1 \,\,\,.\nonumber 
\eea     
The vertex operators are defined by $V_{nm}(x)  = 
:e^{i\alpha_{nm}\Phi(x)}:$ where $\Phi(x)$ is a free scalar 
field and the charges $\alpha_{nm}$ defined by
\EQ
\alpha_{nm} = {1\over 2}(1-n)\alpha_-+{1\over 2}(1-m)\alpha_+~.
\EN
The conformal dimension of the operator $V_{nm}(x)$ is given by 
$
\Delta_{nm} = -\overline{\alpha_{nm}}\alpha_{nm}$ 
with
\EQ
\overline{\alpha_{nm}} = 2\alpha_0-\alpha_{nm} = 
{1\over 2}(1+n)\alpha_-+{1\over 2}(1+m)\alpha_+~.
\EN                                                                          
The integrals encountered in the our computation 
were of two types. 

In the first case, no screening charge is needed, therefore they 
can be computed  in a straightforward way. For instance, this is 
the case of the integral of the $4$-point correlation function 
$\la\vp_1(x_1)\vp_2(x_2)\vp_2(x_3)\vp_3(x_4)\ra$. As an example, 
we provide the calculation of $(C_{22}^3)^{(1)}_1$ 
in the magnetic deformation of the TIM:
$$
(C_{22}^{3(1)})_1 = -\int\limits^{'}\la
\vp_3(\infty)\vp_2(1)\vp_1(z)\vp_2(0)\ra~. $$
Since $2\alpha_{12}+\alpha_{22}+\alpha_{21}=2\alpha_0$, 
no screening charge is needed.
Therefore, 
\EQ
(C_{22}^{3(1)})_1 = N\int d^2 z~ |z|^{-{1\over 5}}|z-1|^{-{1\over 5}} 
\,\,\,,
\EN
where $N$ is a normalization factor which is fixed by the operator
algebra and the structure constants (see Table 2):
$$
N=c_4 c_5 = {1\over 2}\sqrt{ { \Gamma({4\over 5})\Gamma^{3}({4\over 5})
\over \Gamma({1\over 5})\Gamma^{3}({3\over 5}) } } \,\,\,.
$$
Using 
$$
\int d^2 z~ |z|^{2a} |z-1|^{2b} = -S(b) \,B(1+a,1+b)\,
B(1+b,-1-a-b)~,
$$
with 
\[
B(\alpha,\beta) = { \Gamma(\alpha)\Gamma(\beta)\over 
\Gamma(\alpha+\beta) } 
\,\,\,\,\,\,\,,
\,\,\,\,\,\,\,
S(x) \equiv \sin\pi x\,\,\,,
\]
we find:
\EQ
(C_{22}^{3(1)})_1 = - {25\over 32} { S^2({1 \over 10}) 
S^{{1\over 2}}({1\over 5}) \over S^{{3\over 2}}({2 \over 5}) }
{ \Gamma^4({9\over 10})\Gamma({1\over 5}) \over \Gamma^3({3\over 5}) } 
\,\,\,.
\EN

In the second case, one screening operator is needed and the integral 
takes the following form:
\EQ
Z(a,b,c,d,e) = \int d^2w\int d^2 z
|z|^{2a}|1-z|^{2b}|w-z|^{2c}|w|^{2d}|1-w|^{2e} \,\,\,.
\EN
The exponents $a,b,c,d,e$ are computed in the Coulomb gas formalism 
\cite{DF}. Notice that the exponent $d$ is in fact $d'-s/2$, where 
$s$ comes from the Mellin regularization scheme (see Appendix A).
It can be shown using transformations in the complex plane 
that \cite{DF}
\bea
Z(a,b,c,d,e)&= &
S(a+c)^{-1}\int d^2w |w|^{2d} |1-w|^{2e}\times  \nn\\
& &
\left[S(a+b+c)S(b)|I_1(a,b,c;w)|^2+S(a)S(c)|I_2(a,b,c;w)|^2\right]
\,\,\,, 
\nn
\eea
where  
\bea
I_1(a,b,c;\eta)&\equiv&
{\Gamma(-a-b-c-1)\Gamma(b+1)\over\Gamma(-a-c)}
{}_2F_1(-c,-a-b-c-1;-a-c;\eta)\nn\\
I_2(a,b,c;\eta)&\equiv&
\eta^{1+a+c}
{\Gamma(a+1)\Gamma(c+1)\over\Gamma(a+c+2)}
{}_2F_1(-b,a+1;a+c+2;\eta)
\,\,\,.\eea
The generalized hypergeometric functions are defined by 
$${}_pF_q(a_1,\cdots,a_p;b_1,,\cdots,b_q;z)
\equiv\sum_{k=0}^{\infty} {(a_1)_k\cdots (a_p)_k\over k!
 (b_1)_k \cdots (b_q)_k} z^k \,\,\,,$$
with $(a)_k\equiv\Gamma(a+k)/\Gamma(a)$.

There are several ways to compute $Z(a,b,c,d,e)$. We have used
two equivalent methods in order to have a non-trivial check. Let 
us explain briefly them for completeness.

$\bullet$ The first method is the one considered in \cite{GM2}.
Using monodromy properties of the integral,
we can write $Z$ as :
$$Z=z_{11}J_1^2+z_{22}J_2^2+z_{12}J_1J_2~,$$
with 
\bea
J_1 &\equiv& \int\limits_0^1 z^d(1-z)^eI_1(a,b,c,z) = 
B(b+1,-a-b-c-1) \, B(d+1,e+1) \nn\\  
& &\,\,\,\times  
{}_3F_{2}(-c,-a-b-c-1,d+1;-a-c,2+d+e;1) \,\,\, ,
\eea
and
\bea
J_2 &\equiv& \int\limits_0^1 z^d(1-z)^eI_2(a,b,c,z) = 
B(a+1,c+1) \, B(2+a+c+d,e+1) \nn \\
& & \,\,\, \times {}_3F_{2}(-b,a+1,2+a+c+d;a+c+2,3+a+c+d+e;1)\,\,\, .
\eea
In these formulae the $z_{ij}$ are defined by:
\bea
z_{11}& =&
-{1\over 4} S(a+c)^{-2} S^{-1}(c + d + e)S^{-1}(a + b + c + d + e)
\times\nn\\
& &
      S(b)S(a + b + c)S(d) \times\nn\\                   
& &
      ( S(b - c - d) - S(b + c - d) + S(b + c + d) - 
        S(2 a + b + c + d)   \nn\\
& &
 - S(b + c + d + 2 e) + 
        S(2 a + b + 3 c + d + 2 e)) \\
z_{22} &=& {1\over 4}  S(a+c)^{-2}
S^{-1}(c + d + e)S^{-1}(a + b + c + d + e)\times\nn\\
& &
     S(a)S(c)S(a + c + d)\times\nn\\
& & 
    (S(a + b - d) + S(a - b + d) - S(a + b + d) + 
       S(a + b + 2 c + d) \nn\\
& &
 - S(a - b - d - 2 e) - 
       S(a + b + 2 c +  d + 2 e)) \\
z_{12} &  =&  2 S(a+c)^{-2}
S^{-1}(c + d + e)S^{-1}(a + b + c + d + e)
     \times  \nn\\ 
 & &
  S(a)S(b)S(c)S(a + b + c)S(d) S(a + c + d)
\eea

$\bullet$ The second method is the one considered in \cite{DPP}. 
It consists in decomposing $Z$ into its holomorphic and
antiholomorphic parts:
\EQ
\label{int}
I = s(b) s(e) \left[ J_1^+  J_1^- + J_2^+  J_2^-\right]
+ s(b) s(e+c) J_1^+ J_2^- + s(b+c) s(e) J_2^+ J_1^- \,\,\,,
\EN
where
\bea
\label{jis}
J_1^+ &=& J(a,b,c,d,e)~;~J_2^+ = J(b,a,c,e,d)\,\,\,; \nn \\
J_1^- &=& J(b,-2-a-b-c,c,e,-2-d-e-c) \,\,\,;\\
J_2^- &=& J(-2-a-b-c,b,c,-2-d-e-c,e) \,\,\,,\nn
\eea
with the notation
\begin{eqnarray*}
\label{hyp}
&& J(a,b,c,d,e) =
\int\limits_{0}^{1} du \int\limits_{0}^{1} dv~
u^{a+d+c+1} (1-u)^b v^{d}(1-v)^{c} (1-uv)^{e} = B(1+c,1+d) \times\\
&& \times B(2+a+d+c,1+b) {}_3F_2(-e,2+c+d,1+d,3+a+b+c+d,2+c+d;1)\nn.
\end{eqnarray*}
The $J$ integrals appearing in (\ref{int}) are not independent. 
Using a contour deformation it can be shown that they are 
related as 
\EQ
\label{rel1}
S(a+b+c)J_1^- + S(a+b)J_2^- = {S(a)\over S(c+d+e)}
\left( S(d) J_1^+ + S(c+d)J_2^+ \right) \,\,\,;
\EN
\EQ
\label{rel}
S(c+d+e)J_2^- + S(d+e)J_1^- = {S(d)\over S(a+b+c)}
\left( S(a) J_2^+ + S(a+c)J_1^+ \right) \,\,\,.
\EN
This has the advantage that some of the $J_i^{\pm}$ 
can be computed in an easier way than others.

In most cases, we were able to compute exactly these types of 
integrals in terms of a product of Gamma functions 
by using some known relations of hypergeometric functions at argument 
$z=1$ \cite{Prud}. Given their cumbersome expressions, the results 
are not reported here. When no close forms were available, the integrals 
were determined numerically (using the two different representations
above) by directly calculating the hypergeometric functions ${}_3F_2(z)$ 
at $z=1$. In this case, we used fast convergent expressions of the
hypergeometric functions in order to reach rather accurate results 
(with approximately 0.5 \% of confidence).

We have performed the above set of calculations for the magnetic, 
thermal and submagnetic perturbations. The results are in Tables 
\ref{tc1}, \ref{tc2} and \ref{tc3}. We did not pursue this calculation 
for the vacancy density perturbation for the UV difficulties explained 
in the text and also because in this case Form Factors were expected 
to provide a reasonable approximation of the correlators in all 
range of $r$.

\section{Regularization of the VEV's.}
The VEV's of primary operators $\Phi_{l,k}$ in the minimal models 
$\cM_{p,p'}$ perturbed by an integrable relevant operator
$\phi_i$ have been conjectured in \cite{russian2}. They can 
be written as 
\EQ
\langle0_s|\Phi_{l,k}|0_s\rangle = B_{(l,k),i,s}(\xi,\eta) \,
g^{\frac{\Delta_{l,k}}{1-\Delta_i}} \,\,\,,
\label{VEVg1}
\EN
where $s$ labels the different vacua. The prefactor 
$B_{(l,k),i,s}(\xi,\eta)$ can be further decomposed 
as 
\EQ
B_{(l,k),i,s}(\xi,\eta) = b_{(l,k),i,s}(\xi) \, Q_i(\xi,\eta) 
\,\,\,,
\EN
where $b_{(l,k),i,s}(\xi)$ is a simple function for any perturbation 
$\phi_i$ and any vacuum $|0_s\rangle$ whereas $Q_i(\xi,\eta)$ can 
be expressed as 
\EQ
Q_i(\xi,\eta) = e^{I_i(\xi,\eta)} \virg I_i(\xi,\eta)=\int_0^{+\infty} dt
F_i(t;\xi,\eta)\,\,\,.  
\label{I_i}
\EN
In Section 4 we have presented the explicit formulae for the three 
integrable deformations of the TIM. The above integral may diverge for 
some values of the two parameters 
\EQ 
\xi = \frac{p}{p'-p} \virg \eta=(\xi +1)l-\xi k \,\,\,.    
\label{xieta}
\EN
In fact, due to the asymptotic behavior of $F_i(t;\xi,\eta)$ 
\EQ
F_i(t;\xi,\eta)\sim e^{-a_i(\xi,\eta)t} 
\virg t\rightarrow +\infty \,\,\,,
\label{F_i}
\EN
this occurs when $a_i(\xi,\eta)\leq 0$. In this case, 
the integral $I_i(\xi,\eta)$ needs to be regularized in order to extract  
its physical value. Its regularization may be performed in a number 
of equivalent ways, for instance by means of an analytic prolongation 
in $\xi$ (from values for which $Q_i(\xi,\eta)$ converges), possibly 
also using a sufficient number of times the reflection equations 
satisfied by the VEV. Here we present another simple method of controlling 
the divergences of $I_i(\xi,\eta)$, specializing our discussion to the 
VEV $\langle \Phi_{2,1}\rangle_4$ of the sub-leading magnetization 
operator of the TIM in the massive phase reached by the perturbation 
of the vacancy operator $\phi_{1,3} \equiv t\equiv \vp_4$, with $g_4 < 0$. 
This example, somehow, presents all possible types of divergences of 
the above integrals. Let us consider initially the general expression 
of the function $F$ in the case of $\phi_{1,3}$ perturbation \cite{russian2} 
\EQ
F_{1,3}(t;\xi,\eta)=\frac{1}{t}\left(
\frac{\cosh2t\, \sinh(\eta-1)t}{2\cosh t \,
\sinh(\xi t) \sinh(\xi+1)t}-\frac{\eta^2-1}{2\xi(\xi+1)} 
e^{-4t}\right) \,\,\,.
\label{F_{1,3}}
\EN
In our example $\xi=4$ and $\eta=6$, hence  
\EQ
I_{4}(4,6) = \int_0^{+\infty} dt \,F_{4}(t;4,6) \virg
F_{4}(t;4,6) = \frac{1}{t}\left(\frac{\cosh2t\, \sinh7t}
{2\cosh t \,\sinh4t}-\frac{7}{8}
e^{-4t}\right) \,\,\,,
\label{divergentintegral}
\EN
and the asymptotic behavior 
\EQ
F_{4}(t;4,6) \sim e^{4t} 
\virg t\rightarrow +\infty
\label{F_{epsilon'}}
\EN
leads to the divergence of the above integral. 
The complete VEV under investigation can be expressed as 
\EQ
\langle0_s|\sigma'|0_s\rangle = \frac{\sin\frac{3}{2}\pi s}
{\sin\frac{\pi}{4}s}\,
\left(\frac{5}{2}\right)^{\frac{7}{8}}\,
\left[\pi\frac{21}{25}|g_4|\sqrt{\frac{\Gamma(\frac{1}{5})
\Gamma(-\frac{7}{5})}{\Gamma(\frac{4}{5})
\Gamma(\frac{12}{5})}}\right]^{\frac{35}{32}}\,Q_{4}(4,6)\,\,\,,
\label{sigma'VEV}
\EN
where $s = 1,2,3$ labels the three different vacua. Assuming a 
regularization of $Q_{4}(4,6)$, notice that for $s=2$ 
the first term in (\ref{sigma'VEV}) implies 
\EQ
\langle0_s|\sigma'|0_s\rangle = 0 \,\,\,.
\EN
To compute $I_4(4,6)$, let us break the integral as 
\EQ
I_{4}(4,6) = \int_0^1 dt \,F_{4}(t;4,6) + \int_1^{+\infty} dt\,
F_{4}(t;4,6) \,\,\,.
\EN
The first integral is always convergent since, in general
\EQ
\lim_{t\rightarrow 0} F_{1,3}(t;\xi,\eta) = \makebox{\rm const} \,\,\,,
\EN
thanks to a compensation between the two terms in (\ref{F_{1,3}}).
Hence it is sufficient to make an analytic prolongation of the 
second integral by subtracting its divergent part. This can be 
done by expressing initially the second integral as 
\EQ
\int_1^{+\infty} dt F_{4}(t;4,6) = \int_1^{+\infty} 
\left( F_{4}(t;4,6) - \frac{e^{4t}}{2t} + \frac{e^{2t}}{2t}\right)
\,dt + Y(-4)- Y(-2) \,\,\,,
\label{analyticprolongation}
\EN
and then by making an analytic prolongation of the function 
\EQ
Y(a) = \int_1^{+\infty} \frac{dt}{t} e^{-at} \virg \Re e(a) > 0
\,\,\,,
\label{Y}
\EN
to the domain $\Re e(a)\leq 0$. Notice that although the analytic 
extension of the function $Y(a)$ is not monodromic ($a=0$ is in 
fact a branch--cut point), its exponential $e^{Y(a)}$ is however 
uniquely defined and this is precisely the expression which enters 
the formula (\ref{sigma'VEV}) for the VEV. In the punctured complex 
plane $a \in \bC -\{0\}$ we have 
\EQ
e^{Y(a)} = \frac{e^\gamma}{a} e^{-f(a)} \,\,\,,
\label{e^Y}
\EN
where $\gamma=0.577216\dots$ is the Euler-Mascheroni constant and the
holomorphic function $f(a)$ is defined by the power series
\EQ
f(a) = \sum_{n=1}^{\infty} (-1)^n \,\frac{a^n}{n! \,n} \virg a\in\bC 
\,\,\,.
\label{f}
\EN
The last series is fastly convergent and allows good numerical 
estimations. 

The divergences which appear in the expression of the VEV 
proposed in ref.\,\cite{russian2} can be generally tamed as in 
eq.\,(\ref{analyticprolongation}) above, i.e. by subtracting the 
leading (and subleading) exponential terms and then performing 
the analytic continuation of $Y(a)$. However, sometimes 
the first term in the r.h.s. of (\ref{analyticprolongation}), 
i.e. the integral accompanied by the subtractions, is still 
logarithmically divergent. This is in particular the case of 
our example of $\langle \varphi_3 \rangle_4$. When this happens, 
the pure power law behavior in the coupling constant $g_i$ of the VEV 
(\ref{VEVg1}) gets modified. To face this situation, one may 
perform an extra regularization of the integral by making a 
shift of the parameter $\xi$ which characterizes the minimal models
\EQ 
\xi \rightarrow \xi+\epsilon,
\label{shift} 
\EN
and then carefully taking the limit $\epsilon\rightarrow 0$. In 
our example this results in shifting $4 \rightarrow 
4 + \epsilon$, $6 \rightarrow 6 + \epsilon$ and considering the 
expression 
\EQ 
\int_1^{+\infty} 
\left(F_{4}(t;4+\epsilon,6+\epsilon) 
- \frac{e^{4t}}{2t} + \frac{e^{2t}}{2t} - \frac{e^{-\epsilon t}}{t}\right)
\, dt + Y(-4) - Y(-2) + Y(\epsilon) \,\,\,. 
\label{shiftedformula}
\EN
The limit $\epsilon\rightarrow 0$ of the first integral in the previous 
equation is obtained by putting $\epsilon=0$ and hence it can be
calculated by performing a numerical integration. Concerning the 
last term, once used to evaluate the VEV, we have 
\EQ
e^{Y(\epsilon)} = \frac{e^\gamma}{\epsilon} + {\cal O}(1) \,\,\,.
\label{e^Y(epsilon)}
\EN
Let us consider now the dependence of the VEV on the coupling 
constant $g\equiv -g_4$ once the shift (\ref{shift}) has been 
performed. We have  
\EQ
g^{\frac{35}{32}} \rightarrow g^{\frac{35}{32} +
\frac{13}{128} \epsilon} \,\,\,, 
\label{expcorrection}
\EN
since the conformal weights depend on $\xi$. By expanding this 
formula in powers of $\epsilon$ we have 
\EQ
g^{\frac{35}{32}+ \frac{13}{128}
\epsilon} = (g_0)^{\frac{13}{128}\epsilon} g^{\frac{35}{32}}
\left(\frac{g}{g_0}\right)^{\frac{13}{128}\epsilon} =
g^{\frac{35}{32}}\left[ 1 + \frac{13}{128} \epsilon \,\ln\frac{g}{g_0}
+ {\cal O}(\epsilon^2)\right],
\label{expcorrectionexp}
\EN
where we have introduced an arbitrary value of the coupling constant 
$g_0$ for taking the logarithm of the adimensional quantity $\frac{g}{g_0}$. 
Once this expression is multiplied with eq.\,(\ref{e^Y(epsilon)}), 
it gives rise to 
\EQ
g^{\frac{35}{32}} \left(
\frac{e^\gamma}{\epsilon} + \frac{13}{128} e^\gamma
\,\ln\frac{g}{g_0} +
{\cal O}(1) \right) \,\,\,.
\EN
In conclusion, in the example of the VEV (\ref{sigma'VEV}), 
apart from a divergence $\frac{1}{\epsilon}$ which can be 
discarded, it presents a logarithmic part and a
power-law term: 
\EQ
\langle0_s|\sigma'|0_s\rangle =
B_{(2,1),4,s}\, g^{\frac{35}{32}}\,\ln\frac{g}{g_0} + 
C_{(2,1),4,s} \,g^{\frac{35}{32}}. 
\label{logVEV}
\EN
It is important to notice that in the previous expression the constant
$B_{(2,1),4,s}$ is uniquely determined by this procedure and is not 
affected by a change of the reference coupling constant $g_0$. Instead, 
the second constant $C_{(2,1),4,s}$ may be freely modified by rescaling 
the arbitrary value of the reference coupling constant $g_0$.  
In this sense, the constant $B_{(2,1),4,s}$ possesses a precise 
physical value in QFT and it may enter the definitions of amplitude 
ratios as $B$ prefactor.

\section{Form Factors in the thermal sector}
\label{ff}

In this appendix, we will discuss some features of the Form Factors 
in the thermal deformation of the TIM. In particular, we will show 
that it is necessary to determine independently (for instance by a numerical 
method) the one--particle Form Factors of the $\varphi_1$ in order to compute 
its higher particle Form Factors. We will discuss these matrix elements 
in the high--temperature phase of the model with the mass spectrum and 
the $Z_2$ quantum number of the particles given in Table \ref{tspectrum}. 
Let 
\EQ
F_{\al_1,...,\al_n}(\th_1,...,\th_n) = \langle
0|\sigma(0)|A_{\al_1}(\th_1...A_{\al_n}(\th_n)\rangle \,\,\,.
\EN
Since $\sigma(x)$ is a $Z_2$ odd operator, the non--vanishing 
matrix elements will be those on $Z_2$ odd multi--particle states. 
Hence, the matrix elements which contribute to the summation (\ref{defff}) 
are, in increasing order of total energy of the corresponding states:
$F_1$, $F_3$, $F_{12}$, $F_6$, $F_{14}$, $F_{111}$, $F_{23}$, $F_{15},\dots$. 
According to the analysis of ref.\,\cite{DMIMMF,AMV}, the two--particle Form 
Factors can be conveniently written as
\EQ
\label{ff2}
F_{ab}^{\sigma}(\th) = F_{ab}^{min}(\th) \,
{{\cal Q}_{ab}^{\sigma}
\over D_{ab}} \,\,\, ,
\EN
where $F_{ab}^{min}$ is the minimal solution of the set of 
Watson equations which has neither poles neither zero and 
${\cal Q}_{ab}^{\sigma}$ and $D_{ab}$ are polynomials in 
$\cosh\th$. The latter takes into account the pole structure 
of this matrix element (independent of the field) whereas 
the former depends on the field under consideration, in this case 
$\sigma(x)$. To determine the above quantities we need initially 
the expression of the elastic two--particle $S$--matrix of the 
model \cite{MC,FZ} that can be expressed as 
\begin{equation}
\label{smat}
S_{ab}(\theta)=\prod_{\alpha\in {\cal A}_{ab}}
f_{\alpha}(\theta)^{p_{\alpha}}\,\,\,, 
\end{equation}
where 
\EQ  
\label{falpha}
f_{\alpha}(\theta) \equiv 
\frac{\tanh\frac{1}{2}(\theta+i\pi\alpha)}
{\tanh\frac{1}{2}(\theta-i\pi\alpha)}\,\,\,.
\EN
The different amplitudes can be found in Table 2 of 
ref.\,\cite{AMV} and are not reported here. 
The exponents $p_{\alpha}$ denote the multiplicities of the 
corresponding poles (located at $\th=i\pi\alpha$ and 
$\th=i\pi(1-\alpha)$) identified by the indices $\alpha$. 
Correspondingly, $F_{ab}^{\em min}$ is parameterized by:
\EQ
\label{fmin}
F_{ab}^{min}(\th) = \left(-i\sinh\frac{\th}{2}\right)^{\delta_{a,b}}
\prod_{\alpha\in {\cal A}_{ab}} g_{\alpha}(\theta)^{p_{\alpha}} \,\,\, , 
\EN
where $g_{\alpha}(\theta)$ is given by the integral representation 
\EQ
\label{gmin}
g_{\alpha}(\theta) =
\exp\left[2\int_0^{\infty}\frac{d t}{t}
\frac{\cosh\left[(\alpha-1/2)t\right]}{\cosh t/2 \sinh t}
\sin^2(\hat{\th}t/2\pi)\right]\,\,\, ,
\EN
with $\hat{\th} = i\pi-\th$. The polynomials $D_{ab}(\th)$ are 
entirely determined from the poles of the $S$--matrix. According 
to \cite{DMIMMF}, they are given by 
\EQ
D_{ab}(\th)=\prod_{\alpha\in {\cal
A}_{ab}} \Bigl({\cal P}_\alpha(\th)\Bigr)^{i_\alpha}
\Bigl({\cal P}_{1-\alpha}(\th)\Bigr)^{j_\alpha} \,\,\,,
\label{dab}
\EN 
\EQ
\begin{array}{lll}
i_{\alpha} = n+1\,\,\, , & j_{\alpha} = n \,\,\, , &
\mbox{\rm if} \hspace{1cm} p_\alpha=2n+1\,\,\,; \\
i_{\alpha} = n \,\,\, , & j_{\alpha} = n \,\,\, , &
\mbox{\rm if} \hspace{1cm} p_\alpha=2n\,\,\, ,
\end{array}
\EN
where ${\cal A}_{ab}$ and $p_\alpha$ are defined in eq. 
(\ref{smat}) and can be read from Table 2 of ref.\,\cite{AMV}.
The functions
\EQ
\label{pmin}
{\cal P}_{\alpha}(\th) \equiv \frac{\cos\pi\alpha - \cosh\th}
{2 \cos^2\frac{\pi\alpha}{2}}\,\,\, ,
\EN
give a suitable parameterization of the pole of the FF at 
$\th=i\pi\alpha$. The asymptotic behavior of $g_{\alpha}(\th)$ 
and ${\cal P}_{\alpha}(\th)$ is given by 
\EQ
g_{\alpha}(\th) \sim e^{{|\th|\over 2}}~~~~;~~~~
{\cal P}_{\alpha}(\th)\sim  e^{|\th|}~~~~~~~{\rm for}~~\th\to 
\infty \,\,\,.
\EN
An upper bound on the maximal degree of the polynomial 
${\cal Q}_{ab}^{\sigma}$ can be fixed by the constraint 
\cite{DMIMMF} 
\EQ
\label{asscond}
y \leq \D_{\sigma},
\EN
where $y$ is defined by
\EQ
\label{assbe}
\lim_{|\th_i|\to \infty} ~F_{a_1,...,a_n}(\th_1,...,\th_n)
\sim e^{y |\th_i|} \,\,\,.
\EN
Collecting all the above results, let us consider 
the two--particle Form Factor $F_{12}^{\sigma}(\th)$:
\EQ
F_{12}^{\sigma}(\th) =  F_{12}^{min}(\th)
{{\cal Q}_{12}^{\sigma} \over
D_{12}(\th)} \,\,\,.
\EN
By using eqs.\,(\ref{asscond}) and (\ref{assbe}), for the degree
$\delta$ of ${\cal Q}_{12}$ we have $\delta \leq 1$. The residue 
equations (\ref{pole}) at the simple order poles corresponding to 
the bound states supply us with two equations, namely
\EQ
\label{spole}
-i\lim_{\th \rightarrow iu_{ab}^{c}}(\th -iu_{ab}^{c})
F^{\sigma}_{ab}(\th) =
\gamma_{ab}^{c}
F^{\sigma}_{c} \,\, ,
\EN
with $a=1,~b=2,~c=1,3$ and $\gamma_{ab}^{c}$ is given by
\EQ
-i\lim_{\th \rightarrow iu_{ab}^{c}}(\th -iu_{ab}^{c})
S_{ab}(\th)=
\left(\gamma_{ab}^{c}\right)^2~.
\EN
These two equations are able to fix unambiguously $F_{12}^{\sigma}$
provided $F_{1}^{\s}$ and $F_{3}^{\s}$ are known. However, there is 
no way to determine these one--particle Form Factors in the bootstrap
program. Notice, in fact, that the above equations are also 
satisfied by the two--particle FF of the subleading magnetization 
$\sigma'(x)$. Therefore, one needs to extract the one--particle 
FF of $\sigma(x)$ and $\sigma'(x)$ by means of some other 
independent method, as for instance the one provided by the numerical 
Truncated Conformal Space Approach, discussed in Section 5.

\newpage 

{\bf Table Caption}

\vspace{3mm}  
\begin{description}
\item [Table 1]. Kac table of the Tricritical Ising Model. 
\item [Table 2]. Fusion Rules and structure constants of the 
TIM for the scalar fields. 
\item [Table 3]. Operator content and LG fields. 
\item [Table 4]. Nature of QFT's for each 
individual deformation of the TIM. 
\item [Table 5]. Numerical VEV $B_{ij}$ of the four relevant 
operators of the TIM perturbed by the magnetic ($\varphi_1$) operator.
\item [Table 6]. Numerical estimation of the 
one--FF in the magnetic deformation in units of 
$g_1^{\frac{\Delta_i}{1-\Delta_1}}$.
\item [Table 7]. Numerical values of the first correction of 
the structure constants $(C_{ij}^k)^{(1)}_1=-\int ' d^2 z~\la 
\vp_k(\infty)\vp_j(1)\vp_1(z)\vp_i(0)\ra$ where $ 1\leq i,j\leq 4$ 
and $0\leq k\leq 4$ (with the definition $\vp_0\equiv I$, $I$ the
identity operator). The note `2 screening ops.'  means that the
calculation would have required two screening operators in
the Coulomb gas integral.
\item [Table 8]. Estimated amplitudes in the magnetic 
deformation. The number with $^{\dagger}$ refers to the amplitude 
in front of the logarithm. The number with $^*$ refers to 
the finite part of the susceptibility.
\item [Table 9]. Mass spectrum in the high--temperature 
phase.
\item [Table 10]. Numerical and exact (when available) VEV 
$B_{i2}$ of the four relevant operators of the TIM perturbed by 
the thermal ($\varphi_2$) operator.
\item [Table 11]. Numerical estimation of the first four  
one--particle FF in the high temperature phase in units of 
$g_2^{\frac{\Delta_i}{1-\Delta_2}}$. Some of them vanish 
in virtue of the $Z_2$ spin reversal symmetry.
\item [Table 12]. Numerical estimation of the 
first two one--particle FF in the low 
temperature phase in units of $\mid g_2
\mid^{\frac{\Delta_i}{1-\Delta_2}}$.
\item [Table 13]. Numerical values of the first correction 
of the structure constants $(C_{ij}^k)^{(1)}_2=-\int ' d^2 z~\la 
\vp_k(\infty)\vp_j(1)\vp_2(z)\vp_i(0)\ra$ where $ 
1\leq i,j\leq 4$ and $0\leq k\leq 4$ (with the definition 
$\vp_0\equiv I$, $I$ the identity operator).
\item [Table 14]. Estimated amplitudes in the high temperature 
phase. The number with $^*$ refers to the finite part of the
susceptibility.
\item [Table 15]. Estimated amplitudes in the low temperature phase. 
The number with $^{\dagger}$ refers to the exact amplitude in front of the 
logarithm. The number with $^*$ refers to the finite part of the
susceptibility.
\item [Table 16]. Exact VEV of the four relevant operators in 
$\sigma'$ deformation (from \cite{russian2}). The numbers with  
$^\dag$ refer to the amplitudes in front of the logarithm.
\item [Table 17]. Numerical estimation of some one--particle 
FF in the subleading magnetic perturbation in units of 
$g_3^{\frac{\Delta_i}{1-\Delta_3}}$. Those relative to 
$\vp_3$ and $\vp_4$ are not accessible.
\item [Table 18]. Numerical values of the first correction of 
some structure constants 
$(C_{ij}^k)_3^{(1)}$ with $1\leq i,j\leq 4 $ and $0\leq k\leq 4$.
\item [Table 19]. Estimated amplitudes in the subleading 
magnetization deformation for both vacua (the subscripts $a$ and $b$ 
are respectively for $|0_2\ra$ and $|0_4\ra$) obtained by the integral 
of the correlators  and some exact sum rules results. The number 
with $^{\dagger}$ refers to the amplitude in front of the logarithm.
\item [Table 20]. Exact VEV of the four relevant operators  
in $\vp_4\equiv t$ perturbation for $g_4 < 0$ (from \cite{russian2}).
The numbers with $^\dag$ refer to the amplitudes  
in front of the logarithm.   
\item [Table 21]. Estimated amplitudes in the vacancy density 
deformation for the three vacua. The number with $^{\dagger}$ refers 
to the amplitude in front of the logarithm.
\item [Table 22]. Amplitude ratios $R^2_{jk}
= {\Gamma_{jk}^{2+}\over \Gamma_{jk}^{2-}}$. The subscripts $2\pm$ indicate
the high or low temperature phases.
\item [Table 23]. Universal ratios $(Q_2)^i_{jk}$ for 
$i,j,k=1,2^+,2^-$. 
\item [Table 24]. Universal ratios $(R_c)_{jk}^{1}$, $(R_c)_{jk}^{2-}$,
$(R_c)_{jk}^{3a}$ and $(R_c)_{jk}^{3b}$ where $2-$ indicates the 
low temperature phase and $3a,3b$ are used to label the two 
different vacua $|0_2\ra$ and $|0_4\ra$ in the $g_3\vp_3$ 
deformation. The other ratios 
are provided by the sum rule: $(R_c)^j_{ji}={\D_j \D_i\over (1-\D_j)^2}$.
\item [Table 25]. Universal ratios $R_{\xi}^i$ and  
$(R_A)_{j}^{i}$ for $i,j=1,2^-,2^+$. 
\item [Table 26]. Universal ratio $(R_{\chi})_{j}^{i}$ for $i,j=1,2,3$.
We also have $(R_{\chi})_{j}^{j}=-{\D_j\over (1-\D_j)}$ according to 
the sum rules and $(R_{\chi})_{4}^{j}=0$ because $B_{44}=0.$
\end{description}

\newpage                                                               
\vspace{3mm}
\begin{table}[t]
\begin{center}
\begin{tabular}{|ccccc|}\hline
\hspace{1mm} &$3 \over 2$ & $7 \over 16$ & $0$ &\hspace{1mm} \\
\hspace{1mm} & $6 \over 10$ & $3 \over 80$ & $1 \over 10$ 
&\hspace{1mm} \\
\hspace{1mm} &$1 \over 10$ & $3 \over 80$ & $6 \over 10$ 
&\hspace{1mm} \\
\hspace{1mm} &$0$ & $7 \over 16 $ & $3 \over 2$ &\hspace{1mm} \\
\hline
\end{tabular}
\end{center}
\caption{ \label{tkac}} 
\end{table}

\vspace*{1cm }
\begin{table}[h]
\begin{center}
\begin{tabular}{|clc|clc|}\hline
\hspace{1mm} & {\em even} $*$ {\em even} &\hs & \hs & \hs &\hs \\
\hs &$\epsilon*\epsilon=[1]+c_1 \hs [t]$
&\hs &\hs & \hs &\hs \\
\hs & $t * t=[1] +c_2 \hs [t]$ &\hs & \hs
& \hs & \hs\\
\hs &$\epsilon * t =c_1\hs [\epsilon] +c_3\hs [\varepsilon'']$
&\hs &\hs & $c_1=\frac{2}{3}\sqrt{\frac{\Gamma(\frac{4}{5})
\Gamma^3(\frac{2}{5})}{\Gamma(\frac{1}{5})
\Gamma^3(\frac{3}{5})}} $ &\hs\\
\hs &\hs &\hs &\hs &$c_2=c_1$ & \hs \\
\cline{1-3}
\hs & {\em even} $*$ {\em odd} &\hs & \hs & $c_3 ={3\over 7}$ &\hs \\
\hs &
$\epsilon *\sigma'=c_4 \hs [\sigma]$
&\hs &\hs & $c_4={1\over 2} $ &\hs \\
\hs & $\epsilon * \sigma=c_4 \hs [\sigma'] +c_5 \hs [\sigma]$ &\hs 
& \hs & $c_5 = {3\over 2}c_1$ &\hs \\
\hs& $t * \sigma'=c_6 \hs [\sigma]$ &\hs &
\hs & $c_6={3\over 4}$ &\hs \\
\hs & $t *\sigma=c_6 \hs [\sigma']+c_7 \hs [\sigma]$
&\hs & \hs & $c_7 = {1\over 4}c_1$ &\hs \\
\hs &\hs &\hs &\hs & $c_8 = {7\over 8}$ &\hs \\
\cline{1-3}
\hs & {\em odd} $*$ {\em odd} &\hs & \hs & 
$c_9 = {1\over 56}$ &\hs \\
\hs &
$\sigma'*\sigma' = [1] + c_8 \hs [\varepsilon'']$
&\hs &\hs & \hs &\hs \\
\hs & $\sigma'*\sigma=c_4 \hs [\epsilon] +c_6 \hs [t]$
&\hs 
& \hs & \hs &\hs \\
\hs &$\sigma*\sigma=[1]+c_5 \hs [\epsilon]+
c_7 \hs [t]+c_9\hs [\varepsilon'']$ 
&\hs & \hs & \hs & \hs \\
\hs & \hs & \hs & \hs & \hs &\hs\\
\hline
\end{tabular}
\end{center}
\caption{ \label{tfusion}}
\vspace{1cm}
\end{table}

\newpage

\begin{table}[t]
\begin{center}
\begin{tabular}{|ccclcccl|}\hline
\hspace{1mm} & $\sigma$ & = & $[{3 \over 80},{3 \over 80}]$ &=& $
\Phi$ & & magnetization\\
\hspace{1mm} & $\epsilon $ & = & $[{1 \over 10},{1 \over 10}]$ &=& 
$\Phi^2$ & & 
energy\\
\hspace{1mm} & $\sigma'$ & = & $[{7 \over 16},{7 \over 16}]$ &=&$
\Phi^3$ & & sub-magnetization\\
\hspace{1mm}& $t$ & = & $[{6 \over 10},{6 \over 10}]$ &=& 
$ \Phi^4$ & & 
vacancy density \\
\hspace{1mm}& $\varepsilon''$ & = & $[{3 \over 2},{3 \over 2}]$ &=& 
$ \Phi^6 $
& & irrelevant\\
\hline
\end{tabular}
\end{center}
\caption { \label{tope}}
\vspace{15mm}
\end{table}

\vspace{3mm}
\begin{table}[h]
\begin{tabular}{|c|c||c|c||c||c|c||}\hline
& $g_1$ & $g_2^+$ & $g_2^-$ & $g_3$ & $g_4^+$ & $g_4^-$ \\
\hline 
& & & & & &  \\ 
QFT & \mbox{Non-} & \mbox{integrable} & 
\mbox{integrable} & integrable & integrable & integrable \\
&\mbox{Integrable} & $E_7$  & 
$E_7$  & (kinks) & (massless & (Susy \\
& &(high-temp) &(low-temp) && flow) &kinks)  \\ 
\hline
\end{tabular}
\vspace{3mm}
\caption {\label{tqft}}
\vspace{3mm}
\end{table}

\begin{table}[h]
\begin{center}
\begin{tabular}{|c || c| c| c| c||}
\hline
~ & $B_{11}$& $B_{21}$ &$B_{31}$ &$B_{41}$ \\
\hline
\hline
$g_1>0$ (num)& $-1.539(6)$ & $1.35(6)$ & $-1.5(5)$ &$1.9(2)$ \\
\hline
\hline
\end{tabular}
\end{center}
\vspace{3mm}
\caption{\label{tvevphi1}}
\vspace{15mm}
\end{table}

\newpage

\begin{table}[t]
\vspace{5mm}
\centering
\begin{tabular}{|c || c| c| c| c||}
\hline
~ & $\varphi_1$& $\varphi_2$ &$\varphi_3$ &$\varphi_4$ \\
\hline
\hline
$\langle0|\varphi|1\rangle_1$ & $- 0.52(0)$ & $1.1(7)$ 
& $- 4.(9)$ &$7.(4)$ \\
\hline
$\langle0|\varphi|2\rangle_1$& $- 0.2(1)$  & $0.5(6)$ 
& $- 3.(3)$&$ 5.(8)$ \\
\hline
\end{tabular}
\vspace{3mm}
\caption{\label{tff1}}
\vspace{5mm}
\end{table}

\begin{table}[h]
\vspace{3mm}
\begin{center}
\begin{tabular}{|c c c||c c c|}
\hline
$(C_{11}^1)^{(1)}_1$  & = & $2$ screening ops. & $(C_{11}^3)^{(1)}_1$ 
& = & $-0.018583...$
\\ \hline 
$(C_{13}^1)^{(1)}_1$  &$\ap$ & $ -0.482(1)$ & $(C_{13}^3)^{(1)}_1$ & 
$\ap$ & $ 0.395(0)$
\\ \hline 
$(C_{33}^1)^{(1)}_1$  & = & $ -0.214849...$ & $(C_{33}^3)^{(1)}_1$ & 
= & $ 0 $
\\ \hline
$(C_{12}^0)^{(1)}_1$  &= &  $-0.112093...$ & $(C_{12}^2)^{(1)}_1$ & 
$\ap$ & $ 0.517(2)$
\\ \hline
$(C_{12}^4)^{(1)}_1$  &$\ap$ &  $-0.015(0)$ & $(C_{14}^0)^{(1)}_1$ & = 
& $-2.548155... $
\\ \hline
$(C_{14}^2)^{(1)}_1$  &$\ap$ &  $0.260(1)$ & $(C_{14}^4)^{(1)}_1$ & = 
& $2$ screening ops.
\\ \hline
$(C_{23}^0)^{(1)}_1$  &$=$ &  $-2.816773...$ & $(C_{23}^2)^{(1)}_1$ 
& = &$0.683830...$
\\ \hline
$(C_{23}^4)^{(1)}_1$  &$=$ &  $0.3787045...$ & $(C_{34}^0)^{(1)}_1$ 
& = &$-0.922183...$
\\ \hline
$(C_{34}^2)^{(1)}_1$  &$=$ &  $0.259270...$ & $(C_{34}^4)^{(1)}_1$ 
& = &$-0.665160...$
\\ \hline
$(C_{22}^1)^{(1)}_1$  &$\ap$ &  $0.266(0)$ & $(C_{22}^3)^{(1)}_1$ 
& = &$-2.215418...$
\\ \hline
$(C_{24}^1)^{(1)}_1$  &$\ap$ &  $-1.54(9)$ & $(C_{24}^3)^{(1)}_1$ 
& = &$0.504471...$
\\ \hline
$(C_{44}^1)^{(1)}_1$  &= & $2$ screening ops. & $(C_{44}^3)^{(1)}_1$ & = &$0$
\\ \hline
\end{tabular}
\end{center} 
\caption{\label{tc1}}
\vspace{15mm}
\end{table}



\begin{table}[t]
\begin{center}
\begin{tabular}{|c|c|c|c|}
\hline
\hspace{-10pt}
\begin{tabular}{c}Susceptibilities \end{tabular}
\hspace{-10pt}
&
\begin{tabular}{c} Integration \end{tabular}
\hspace{-10pt}
&
\begin{tabular}{c} TCSA  \end{tabular}
\hspace{-10pt}
&
\begin{tabular}{c} Sum Rule\end{tabular}
\hspace{-10pt}
\\
\hline
$\Gamma_{11}^1$&0.05(7) &0.059(6)&0.06 
\hspace{-10pt}
\\
\hline
$\Gamma_{12}^1$&-0.13(6) &-0.139(7)&-0.1396 
\hspace{-10pt}
\\
\hline
$\Gamma_{13}^1$ &0.69(6)&0.68(7) &0.70 
\hspace{-10pt}
\\
\hline
$\Gamma_{14}^1$ &-1.2(1)&-1.1(4) &-1.2(0)
\hspace{-10pt}
\\
\hline
$\Gamma_{22}^1$ &0.31(7)&0.32(7) &
\hspace{-10pt}
\\
\hline
$\Gamma_{23}^1$ &-1.7(3)&-1.6(7) &
\hspace{-10pt}
\\
\hline
$\Gamma_{24}^1$ &3.(0)&2.(8)&
\hspace{-10pt}
\\
\hline
$\Gamma_{33}^1$ &15.(3)& &
\hspace{-10pt}
\\
\hline
$\Gamma_{34}^1$ &$3.76(9)^{\dagger}$ & &
\hspace{-10pt}
\\
\hline
$\Gamma_{44}^1$ &$-15.(5)^*$ & &
\hspace{-10pt}
\\
\hline
\end{tabular}
\end{center}
\caption{\label{tchi1}}
\vspace{3mm}
\end{table}

\begin{table}[h]
\vspace{3mm}
\begin{center}
\begin{tabular}{|cclc|l|l|} \hline
$m_1$ &=& $M$ & & 1  & \hspace{1mm} odd \\
$m_2$ &=& $2 M \cos({5\pi \over 18})$ & & 1.28557 & \hspace{1mm} even \\
$m_3$ &=& $2 M \cos({\pi \over 9})$ & & 1.87938 & \hspace{1mm} odd \\
$m_4$ &=& $2 M \cos({\pi \over 18})$ & & 1.96961 & \hspace{1mm} even \\
$m_5$ &=& $4 M \cos({\pi \over 18}) \cos({\pi \over 9})$ & & 2.53208 & 
\hspace{1mm} even \\
$m_6$ &=& $4 M \cos({2\pi\over 9})\cos({\pi \over 9}) $ & & 2.87938 &
\hspace{1mm} odd \\
$m_7$ &=& $4 M \cos({\pi \over 18}) \cos({\pi \over 9})$ & & 3.70166 & 
\hspace{3mm} even\\
\hline
\end{tabular}
\end{center}
\caption{\label{tspectrum}}
\vspace{3mm}
\end{table}

\newpage

\newpage
\begin{table}[t]
\begin{center}
\begin{tabular}{|c || c| c| c| c||}
\hline
~ & $B_{12}$& $B_{22}$ &$B_{32}$ &$B_{42}$ \\
\hline
\hline
$g_2>0$ (num)& $0$ & $-1.46(6)$ & $0$ &$3.(4)$ \\
\hline
$g_2>0$ (exact)& $0$ & $-1.46839\dots$ & $0$ &$3.70708\dots$ \\
\hline
$g_2<0$  (num)& $\pm 1.59(0)$ & $1.46(6)$ & $\pm 2.3(8)$ &$3.(5)$ \\
\hline
$g_2<0$ (exact)& $\pm 1.59427\dots$ & $1.46839$ & $\pm 2.45205 
\dots$ &$3.70708\dots$ \\
\hline
\end{tabular}
\end{center}
\vspace{3mm}
\caption{\label{tvev}}
\end{table}

\begin{table}[h]
\centering
\begin{tabular}{|c || c| c| c| c||}
\hline
~ & $\varphi_1$& $\varphi_2$ &$\varphi_3$ &$\varphi_4$ \\
\hline
\hline
$\langle0|\varphi_i|1\rangle_{2+}$ & $ 0.78(2)$ & $0$ & $- 6.(4)$ &$0$ \\
\hline
$\langle0|\varphi_i|2\rangle_{2+}$ & $0$  & $ 1.1(9)$ & 0 &$- 11.(1)$ \\
\hline
$\langle0|\varphi_i|3\rangle_{2+}$ & $ 0.2(4)$ & $0$ & $- 4.(3)$ &$0$ \\
\hline
$\langle0|\varphi_i|4\rangle_{2+}$ & $0$  & $ 0.5(9)$ & $0$ &$- 8.(7)$ \\
\hline
\end{tabular}
\vspace{3mm}
\caption{\label{tff2}}
\vspace{3mm}
\end{table}

\newpage

\begin{table}[t]
\centering
\begin{tabular}{|c || c| c| c| c||}
\hline
~ & $\varphi_1$& $\varphi_2$ &$\varphi_3$ &$\varphi_4$ \\
\hline
\hline
$\langle0|\varphi_i|1\rangle_{2-}$ & $0.50(5) $  & $ 1.1(9)$ &$6.3(5)$  &$ 11.(1)$ \\
\hline
$\langle0|\varphi_i|2\rangle_{2-}$ & $0$  & $ 0.5(9)$ &$0$  &$ 8.(7)$ \\
\hline
\end{tabular}
\vspace{3mm}
\caption{\label{tff2m}}
\vspace{3mm}
\end{table}


\vspace*{5mm}
\begin{table}[h]
\begin{center}
\begin{tabular}{|c c c||c c c|}
\hline
$(C_{11}^0)^{(1)}_2$  & = &$0.223579...$ & $(C_{11}^2)^{(1)}_2$ 
& = & $0.266530...$
\\ \hline 
$(C_{11}^4)^{(1)}_2(mr)$ & $\ap$ & $0.1510..(\ln(mr)+b_1(m))$ 
& $(C_{13}^0)^{(1)}_2$ & $=$ & $-2.007437$
\\ \hline 
$(C_{13}^2)^{(1)}_2$  & = & $ 0.677665\dots$ & $(C_{13}^4)^{(1)}_2$ 
& = & $ 0.181313\dots$
\\ \hline
$(C_{33}^0)^{(1)}_2$  &= &  $0$ & $(C_{33}^2)^{(1)}_2$ & $=$ & $ 0$
\\ \hline
$(C_{33}^4)^{(1)}_2(mr)$  &$=$ &  $2.3561..(\ln(mr)+b_2(m)) $ 
& $(C_{12}^1)^{(1)}_2$ &$\ap$ & $0.230(2) $
\\ \hline
$(C_{12}^3)^{(1)}_2$  &$=$ &  $0.109817...$ &  $(C_{23}^1)^{(1)}_2$ & 
= &$-3.766576...$
\\ \hline
$(C_{23}^3)^{(1)}_2$  &$=$ &  $1.840967...$ & $(C_{14}^1)^{(1)}_2$ 
&$\ap$& -3.79(9)
\\ \hline
$(C_{14}^3)^{(1)}_2$  &$=$ &  $1.029533...$ & $(C_{34}^1)^{(1)}_2$ 
& = &$-1.412466...$
\\ \hline
$(C_{34}^4)^{(1)}_2$  &$=$ &  $0.545471...$ & $(C_{22}^2)^{(1)}_2$ 
& = &$1.008826...$
\\ \hline
$(C_{22}^4)^{(1)}_2$  &$=$ &  $0$ & $(C_{24}^0)^{(1)}_2$ & $\ap$ & $-4.19(0)$
\\ \hline
$(C_{24}^2)^{(1)}_2$  &$=$ &  $0$ & $(C_{24}^4)^{(1)}_2(mr)$ & $\ap$ 
&  $2.84(0)(\ln(mr)+b_3(m))$
\\ \hline
$(C_{44}^2)^{(1)}_2$  &$\ap$ &  $0.222(1)$ & $(C_{44}^4)^{(1)}_2$ & = &$0$
\\ \hline
\end{tabular}
\end{center} 
\caption{\label{tc2}}
\vspace{3mm}
\end{table}

\begin{table}[h]
\begin{center}
\begin{tabular}{|c|c|c|c|}
\hline
\hspace{-10pt}
\begin{tabular}{c}Susceptibilities \end{tabular}
\hspace{-10pt}
&
\begin{tabular}{c} Integration \end{tabular}
\hspace{-10pt}
&
\begin{tabular}{c} TCSA \end{tabular}
\hspace{-10pt}
&
\begin{tabular}{c} Sum Rule\end{tabular}
\hspace{-10pt}
\\
\hline
$\Gamma_{11}^{2+}$&0.093(9) &0.093(7)& 
\hspace{-10pt}
\\
\hline
$\Gamma_{12}^{2+}$&0 &0&0 
\hspace{-10pt}
\\
\hline
$\Gamma_{13}^{2+}$ &-0.8(9)&-0.8(8) & 
\hspace{-10pt}
\\
\hline
$\Gamma_{14}^{2+}$ &0&0&0 
\hspace{-10pt}
\\
\hline
$\Gamma_{22}^{2+}$ &0.15(8)&0.16(0) &0.16315...
\hspace{-10pt}
\\
\hline
$\Gamma_{23}^{2+}$ &0&0 &0
\hspace{-10pt}
\\
\hline
$\Gamma_{24}^{2+}$ &-2.(2)&-2.(1)&-2.466...
\hspace{-10pt}
\\
\hline
$\Gamma_{33}^{2+}$ &16.(5)& &
\hspace{-10pt}
\\
\hline
$\Gamma_{34}^{2+}$ &0&0 &0
\hspace{-10pt}
\\
\hline
$\Gamma_{44}^{2+}$ &$-17.(5)^*$ & &
\hspace{-10pt}
\\
\hline
\end{tabular}
\end{center}
\vspace{3mm}
\caption{\label{tchi2}} 
\vspace{3mm}
\end{table}


\begin{table}[h]
\begin{center}
\begin{tabular}{|c|c|c|c|}
\hline
\hspace{-10pt}
\begin{tabular}{c}Susceptibilities \end{tabular}
\hspace{-10pt}
&
\begin{tabular}{c}Integration \end{tabular}
\hspace{-10pt}
&
\begin{tabular}{c} TCSA  \end{tabular}
\hspace{-10pt}
&
\begin{tabular}{c}Sum Rule\end{tabular}
\hspace{-10pt}
\\
\hline
$\Gamma_{11}^{2-}$&0.026(2)&0.026(7)& 
\hspace{-10pt}
\\
\hline
$\Gamma_{12}^{2-}$ &$\pm$0.06(3) &$\pm$0.06(6)&$\pm$0.0662... 
\hspace{-10pt}
\\
\hline
$\Gamma_{13}^{2-}$ &0.4(4)&0.4(2) & 
\hspace{-10pt}
\\
\hline
$\Gamma_{14}^{2-}$ &$\pm$0.8(8)&$\pm$0.8(1) & 
\hspace{-10pt}
\\
\hline
$\Gamma_{22}^{2-}$ &0.15(8)&0.16(1)&0.16315...
\hspace{-10pt}
\\
\hline
$\Gamma_{23}^{2-}$ &$\pm$1.1(2)&$\pm$1.1(0) &$\pm$1.1145...
\hspace{-10pt}
\\
\hline
$\Gamma_{24}^{2-}$ &2.(2)&2.(1)&2.466...
\hspace{-10pt}
\\
\hline
$\Gamma_{33}^{2-}$ &12.(6)& &
\hspace{-10pt}
\\
\hline
$\Gamma_{34}^{2-}$ &$\pm 4.17378...^{\dagger}$ & &
\hspace{-10pt}
\\
\hline
$\Gamma_{44}^{2-}$ &$-17.(5)^*$ & &
\hspace{-10pt}
\\
\hline
\end{tabular}
\end{center}
\vspace{3mm}
\caption{\label{tchi2m}}
\vspace{5mm}
\end{table}


\begin{table}[h]
\centering
\begin{tabular}{|c||c|c|}
\hline
&$B_{i3}$ for $|0\ra=|0_2\ra$ & $B_{i3}$ for $|0\ra=|0_4\ra$ \\ \hline 
\hline
$i=1$& $0.68656\dots$ & $-1.79745\dots$ \\ \hline 
$i=2$& $-0.78093\dots$ & $2.04451\dots$ \\ \hline 
$i=3$& $-17.941605\dots$ & $-17.941605\dots$ \\ \hline 
$i=4$& $2.69611\dots^\dag$ & $-7.05856\dots^\dag$ \\ \hline 
\end{tabular}
\vspace{3mm}
\caption{\label{tvev3}}
\vspace{8mm}
\end{table}

\newpage

\begin{table}[t]
\vspace{3mm}
\centering
\begin{tabular}{|c || c|c ||}
\hline
~ & $\varphi_1$& $\varphi_2$   \\
\hline
\hline
$\langle0_2|\varphi_i|1\rangle_{3}$ & $- 0.42(5) $  & $ 0.9(4)$   \\
\hline
$\langle 0_4|\varphi_i|2\rangle_{3}$ & $ 0.87(2)$  & $ 1.8(3)$  \\
\hline
\end{tabular}
\vspace{3mm}
\caption{\label{tff3}}
\vspace{3mm}
\end{table}

\begin{table}[h]
\begin{center}
\begin{tabular}{|c c c||c c c|}
\hline
$(C_{11}^1)^{(1)}_3(mr)$  & = & $0.7189..(\ln(mr)+C_1(m)) 
$ & $(C_{11}^3)^{(1)}_3$ &$\ap$ & $-0.040(1)$
\\ \hline 
$(C_{12}^0)^{(1)}_3$ & $=$ & $8.79920$ & $(C_{12}^2)^{(1)}_3$ & 
$=$ & $-0.6571$
\\ \hline 
$(C_{12}^4)^{(1)}_3$ & = &0.149171  & $(C_{22}^1)^{(1)}_3(mr)$ 
&$=$ & $ 2.8759..\ln(mr+C_2(m)) $
\\ \hline
$(C_{22}^3)^{(1)}_3$  &= &  $0$ & $(C_{14}^0)^{(1)}_3(mr)$ & 
$=$ & $-4.7123..(\ln(mr)+C_3(m)) $
\\  \hline
\end{tabular}
\end{center} 
\vspace{3mm}
\caption{\label{tc3}}
\vspace{1mm}
\end{table}

\begin{table}[h]
\begin{center}
\begin{tabular}{|c c c||c c c|}\hline
& $\Ga_{ij}^{3a}$&  & &$\Ga_{ij}^{3b}$ &\\ \hline
$\Ga_{11}^{3a}$&$=$&$0.014(3)$ & $\Ga_{11}^{3b}$&$
=$&$0.063(4)$ \\ \hline
$\Ga_{12}^{3a}$&$=$&$-0.03(2)$ & $\Ga_{11}^{3b}$&$
=$&$0.13(7)$\\ \hline
$\Ga_{22}^{3a}$&$=$&$0.076(5)$ & $\Ga_{22}^{3b}$&$
=$&$0.29(0)$\\ \hline
$\Ga_{13}^{3a}$&=&$-0.045770...$ & $\Ga_{13}^{3b}$&=&
$0.11983...$\\ \hline
$\Ga_{23}^{3a}$&=&$0.138832...$ & $\Ga_{23}^{3b}$&=&
$-0.36346...$\\ \hline
$\Ga_{33}^{3a}$&=&$13.954582...$ & $\Ga_{33}^{3b}$&=&
$13.954582...$\\ \hline
$\Ga_{34}^{3a}$&=&$-2.875855...^{\dagger}$ & $\Ga_{34}^{3b}
$&=&$7.52914...^{\dagger}$\\ \hline
\end{tabular}
\end{center}
\vspace{3mm}
\caption{\label{tchi3}}
\vspace{3mm}
\end{table}

\newpage

\begin{table}[h]
\centering
\begin{tabular}{|c||c|c|c|}
\hline
&$ B_{i4}$ for $|0\ra=|0_{-1}\ra$  &$ B_{i4}$ for $|0\ra=|0_{0}\ra$   
&$ B_{i4}$ for $|0\ra=|0_{+1}\ra$  \\ 
\hline \hline
$i=1$& $-1.975669\dots$ & $0$& $1.975669\dots$ \\ \hline 
$i=2$& $2.668319\dots$ & $-2.668319\dots$ & $2.668319\dots$ \\ \hline 
$i=3$& $-10.640138\dots^\dag$ & $0$ & $10.640138\dots^\dag$ \\ \hline 
$i=4$& $0$ & $0$ & $0$ \\ \hline 
\end{tabular}
\vspace{3mm}
\caption{\label{tvev4}}
\vspace{8mm}
\end{table}

\newpage

\begin{table}[t]
\vspace{5mm}
\begin{center}
\begin{tabular}{|c c c||c c c|}\hline
& $(\Ga_{ij}^{4-})_0$&  & &$(\Ga_{ij}^{4-})_{\pm 1}$ &\\ \hline
$(\Ga_{11}^{4-})_0$&$=$&$0.00(5)$ & $(\Ga_{11}^{4-})_{\pm 1}$&$=$&
$0.00(5)$ \\ \hline
$(\Ga_{12}^{4-})_0$&$=$&$0$ & $(\Ga_{12}^{4-})_{\pm 1}$&$=$&$0.16(0)$
\\ \hline
$(\Ga_{22}^{4-})_0$&$=$&$4.4(9)~10^{-3}$ & $(\Ga_{22}^{4-})_{\pm 1}
$&$=$&$4.4(9)~10^{-3}$\\ \hline
$(\Ga_{14}^{4-})_0$&=&$0$ & $(\Ga_{14}^{4-})_{\pm 1}$&=&
$\mp0.1852198\dots$\\ \hline
$(\Ga_{24}^{4-})_0$&=&$0.6670799...$ & $\Ga_{24}^{4-})_{\pm 1}
$&=&$-0.6670799...$\\ \hline
$(\Ga_{34}^{4-})_0$&=&$0$ & $(\Ga_{34}^{4-})_{\pm 1}$&=&$
\mp 11.637651\dots^{\dagger}$\\ \hline
$(\Ga_{44}^{4-})_0$&=&$0$  & $(\Ga_{44}^{4-})_{\pm 1}$&=&$0$\\ \hline
\end{tabular}
\end{center}
\vspace{3mm}
\caption{\label{tchi4}}
\vspace{18mm}
\end{table}

\newpage

\begin{table}[t]
\begin{center}
\begin{tabular}{|ccc||ccc|}
\hline
$R^2_{11} $   &  = &$3.5(4)$  &
$R^2_{13} $   &  = &$-2.0(6)$  \\
\hline
$R_{22}^2 $   &  = &$1$  &
$R_{24}^2 $   &  = &$-1$  \\
\hline
$R_{33}^2 $   & = &$ 1.3(0)$  &
$R_{44}^2$   & = &$1$  \\
\hline
\end{tabular}
\end{center}
\vspace{3mm}
\caption{\label{tr2}}
\end{table}

\begin{table}[h]
\begin{center} 
\begin{tabular}{|ccl||ccl|} \hline 
$(Q_2)^1_{2^+1}$   & = & $ 1.26(0)$     & $(Q_2)^1_{2^-1}$   
& = &  $1.88(4)$ \\ 
\hline 
$(Q_2)^1_{2^+2^+}$ & = & $ 1.97(3)$     & $(Q_2)^1_{2^+2^-}$      
& = &  $1.32(0)$ \\ 
\hline  
$(Q_2)^{2+}_{11}$  & = & $ 1.5(6) $    & $(Q_2)^{2-}_{11}$  
& = &  $0.44(2)$  \\ 
\hline 
$(Q_2)^{2+}_{12^-}$ & = & $ 1.7(0)$     &  &   & \\ 
\hline 
\end{tabular} 
\end{center} 
\caption{\label{trq2}}
\end{table}

\begin{table}[t]
\begin{center} 
\begin{tabular}{|ccl||ccl|} \hline 
$(R_c)_{22}^1$   & = & $ 1.0(5)~10^{-2}$   &  $(R_c)_{23}^1$     
& = & $4.8(5)~10^{-2}$  \\ 
\hline 
$(R_c)_{24}^1$   & = & $ 6.(7)~10^{-2} $   &  $(R_c)_{33}^1$     
& = & $3.(8)~10^{-1} $ \\ 
\hline 
$(R_c)_{34}^1$   & = & $ -7.(6)~10^{-2} $   &  $(R_c)_{44}^1$     
& = & $-2.(5)~10^{-1} $ \\ 
\hline \hline 
$(R_c)_{11}^{2-}$ & = &$ 1.7(0)~10^{-3}$  & $(R_c)_{14}^{2-}$ & = &  
$2.3(3)~10^{-2}$  \\ 
\hline 
$(R_c)_{13}^{2-}$ & = &$ 1.7(9)~10^{-2}$ & $(R_c)_{33}^{2-}$ &  = &  
$3.(4)~10^{-1}$  \\ 
\hline 
$(R_c)_{34}^{2-}$ & = &$ 7.4912...~10^{-2}$ & $(R_c)_{44}^{2-}$ &  = &  
$-2.(0)~10^{-1}$  \\ 
\hline \hline 
$(R_c)_{11}^{3a}$ & = &$ 4.2(3)~10^{-1}$  & $(R_c)_{11}^{3b}$ & = &  
$2.7(3)~10^{-1}$  \\ 
\hline 
$(R_c)_{12}^{3a}$ & = &$ 8.3(2)~10^{-1}$ & $(R_c)_{12}^{3b}$ &  = &  
$5.2(0)~10^{-1}$  \\ 
\hline 
$(R_c)_{22}^{3a}$ & = &$ 1.7(5)$ & $(R_c)_{22}^{3b}$ &  = &  
$9.6(8)~10^{-1}$  \\ 
\hline 
\end{tabular} 
\end{center} 
\vspace{3mm}
\caption{\label{trc}}
\end{table}


\newpage
\begin{table}[t]
\begin{center} 
\begin{tabular}{|ccl||ccl|} \hline 
$R_{\xi}^1$  & = & $ 7.55(7)~10^{-2}$ &  & & \\ 
\hline 
$R_{\xi}^{2+}$   & = & $ 1.07(8)~10^{-1} $ &  $R_{\xi}^{2-}$     
& = & $8.38(9)~10^{-2} $ \\ 
\hline \hline 
$(R_A)_{2+}^{1}$ & = & $0 $  & $(R_A)_{2-}^{1}$ & = &  
$3.91(8)~10^{-2}$  \\ 
\hline 
$(R_A)^{2+}_{1}$ & = & $ 2.95(8)~10^{-1}$ & $(R_A)^{2-}_{1}$ &  = &  
$8.26(0)~10^{-1}$  \\ 
\hline 
\end{tabular} 
\end{center} 
\vspace{3mm}
\caption{\label{trxi}}
\end{table}

\begin{table}[t]
\begin{center} 
\begin{tabular}{|ccl||ccl|} \hline 
$(R_{\chi})_{2}^1$   & = &  $1.1(9)~10^{-1}$ & $(R_{\chi})_3^{1}  
$  & = & $4.2(5)~10^{-1}$  \\ 
\hline 
$(R_{\chi})_{1}^{2-}$   & = &  $4.(0)~10^{-2}$ & $(R_{\chi})_{3}^{2-}  
$  & = & $4.(0)~10^{-1}$  \\ 
\hline 
$(R_{\chi})_{1}^{3a}$ & = & $ 2.(0)~10^{-11}$ & $(R_{\chi})_{1}^{3b}  
$ & =  & $1.8(7)$  \\ 
\hline
$(R_{\chi})_{2}^{3a}$ & = & $ 3.(3)~10^{-4}$  & $(R_{\chi})_{2}^{3b}  
$ & =  & $2.7(8)$ \\ 
\hline 
\end{tabular} 
\end{center} 
\vspace{3mm}
\caption{\label{tchi}}
\end{table}

\end{document}